\documentclass{aastex631}
\usepackage{amsmath}
\usepackage{graphicx}
\usepackage{threeparttable}
\usepackage{txfonts}
\usepackage{soul}

\begin{document}
%\title{A universal relation for non-repeating fast radio bursts, repeating fast radio bursts and radio bursts from galactic magnetar}
%\title{The Spin Period Search of Four Active Repeating Fast Radio Bursts}
\title{False Periodicities in Repeating Fast Radio Bursts Caused by Burst Clustering}

\correspondingauthor{Fa-Yin Wang}
\email{fayinwang@nju.edu.cn}

\author[0009-0006-4945-4734]{Hao-Tian Lan}
\affiliation{School of Astronomy and Space Science, Nanjing University, Nanjing 210023, China}

\author[0000-0001-6021-5933]{Qin Wu}
\affiliation{International Centre for Radio Astronomy Research, Curtin University, Bentley, WA, Australia}
\affiliation{School of Astronomy and Space Science, Nanjing University, Nanjing 210023, China}

\author[0000-0002-9642-9682]{Jia-Wei Luo}
\affiliation{College of Physics and Hebei Key Laboratory of Photophysics Research and Application, Hebei Normal University, Shijiazhuang, Hebei 050024, China}
\affiliation{Shijiazhuang Key Laboratory of Astronomy and Space Science, Hebei Normal University, Shijiazhuang, Hebei 050024, China}
%\email{ljw@hebtu.edu.cn}

\author[0000-0003-4157-7714]{Fa-Yin Wang}
\affiliation{School of Astronomy and Space Science, Nanjing University, Nanjing 210023, China}
\affiliation{Key Laboratory of Modern Astronomy and Astrophysics (Nanjing University) Ministry of Education, China}

\author[0000-0002-7835-8585]{Zi-Gao Dai}
\affiliation{Department of Astronomy, University of Science and Technology of China, Hefei 230026, China}

\begin{abstract} 
Fast radio bursts (FRBs) are millisecond-duration radio transients with extremely high brightness temperatures, but their physical origin remains uncertain. The association of FRB 200428 with a Galactic magnetar provides compelling evidence that at least some FRBs originate from magnetar-like systems. Nevertheless, periodic modulation associated with neutron-star rotation has not been robustly detected in extragalactic FRBs. Motivated by the reported 1.7 s periodicity in the active repeater FRB 20201124A, we conduct a systematic search for short-timescale periodicity in four highly active repeating FRB sources. We employ a sliding-window method that analyzes temporally adjacent bursts within restricted time intervals, allowing transient or intermittently detectable periodic signals to be identified. The search yields numerous apparent period candidates with nominal Poisson-calibrated significances comparable to that reported for the 1.7 s signal, although the inferred periods vary substantially among sources and observing windows. We then perform a nonparametric waiting-time randomization test, in which the observed waiting-time distribution is preserved while the temporal ordering is destroyed. The randomized sequences reproduce the main candidate-period structures seen in the data, demonstrating that these apparent periodicities can arise from the non-Poissonian waiting-time statistics rather than from a coherent periodic process. Rotation-modulated epidemic-type aftershock sequence simulations provide a complementary parametric illustration of the same effect. We therefore find no robust evidence for a short intrinsic periodicity in these data, and the reported 1.7 s signal is not uniquely distinguished from clustering-induced candidates. Robust periodicity searches must evaluate significance against null models that account for the observed waiting-time statistics and the full search procedure.
\end{abstract}

%   Main}
\section{Introduction}\label{sec1}
Fast radio bursts (FRBs) are bright, millisecond-duration radio transients that are believed to originate from extragalactic sources \citep{Cordes2019, Zhang2023, Wang2026}. They were first found in 2007 \citep{Lorimer2007}. Since then, more than 800 FRBs have been discovered \citep{CHIME/FRBCollaboration2019, Chime/FrbCollaboration2023}, but their physical origin still remains one of the most compelling open questions in astrophysics. A wide range of theoretical models have been proposed to explain the physical origin of FRBs \citep{Platts2019}. 
%generally falling into two broad categories: "close-in" models, in which FRBs are generated near or within the magnetosphere of a neutron star \citep{Yang2018, Kumar2020, Lu2020}, and "far-away" models, where the emission originates at greater distances from the central engine \citep{Beloborodov2017, Metzger2019}. 
Among the various progenitor candidates, compact objects, particularly magnetars, have emerged as the leading contenders \citep{Metzger2017,Platts2019, Kumar2020}. In 2020, FRB 20200428 was localized in the direction of SGR 1935+2154 in the Milky Way \citep{CHIME/FRBCollaboration2020b,Bochenek2020}. This observation strongly supports the hypothesis that FRBs are generated by magnetars. 

The periodicity of FRBs is of great significance in understanding their origin and physical nature. 
Using the time of arrival (TOA) of the burst, FRB 20180916B and FRB 20121102A have been found to exhibit periodic activity, with periods of 16.3 days and 157 days, respectively \citep{Chime/FrbCollaboration2020,Rajwade2020,Cruces2021}. Several physical models have been proposed to explain the detected periodicities, the binary orbital model \citep{Ioka2020,Lyutikov2020,Gu2020,Wada2021}, free precession model \citep{Levin2020,Zanazzi2020}, forced precession model \citep{Yang2020}, and the slow-rotation magnetar scenario \citep{Beniamini2020,Lan2024}. Further observations, particularly studies of the period's dependence on observing frequency, dispersion measure (DM), and polarization angle (PA), as well as constraints on the period derivative have placed meaningful limits on the physical origin of FRB 20180916B and have greatly contributed to our understanding of the origin of FRBs \citep{Pastor-Marazuela2021,Katz2021,Li2021,Lan2024,Bethapudi2025}. In addition, FRB 20201124A shows a 26 day rotation measure (RM) periodicity, which could be well fitted by the binary model \citep{Wang2022,Rajwade2023,Xu2025,Wang2025}. Similarly, FRB 20220529 exhibits a possible RM periodicity of approximately 200 days, with its RM evolution over the cycle consistent with predictions from binary system models \citep{Liang2025}. These two observations provide strong evidence supporting the scenario in which at least some repeating FRBs originate in binary systems \citep{Zhao2023,ZhangB2025}.

In addition to long-term periodicities, considerable effort has also been devoted to searching for the short spin periods predicted by magnetar-based FRB models. For example, a Fast Fourier Transform (FFT) periodic search with PRESTO \citep{Ransom2011} and Fast Folding Algorithm (FFA) \citep{Morello2020} were performed on the bursts of FRB 20121102A detected by MeerKAT \citep{Caleb2020}. \cite{ZhangML2018} used a machine-learning pipeline to identify additional bursts from FRB 20121102A and subsequently applied the Rayleigh test to the burst time-of-arrivals to search for possible periodic modulation. \cite{LiD2021} performed a period search with the Lomb-Scargle Periodogram (LSP) methods \citep{Lomb1976,Scargle1982,VanderPlas2018} on the bursts of FRB 20121102A detected by FAST. \cite{Zhang2023} presented 1076 bursts from FRB 20220912A detected by FAST and then, they used LSP and phase-folding method to search the short period ranging from 1 ms to 100 ms. Subsequently, a systematic search for periodicities was carried out on FRB 20201124A and FRB 20240114A using FAST observations with methods employed including FFT, FFA, LSP, as well as multi-component analysis \citep{Niu2022,Zhou2025}. However, despite theoretical expectations that short periodicities in FRBs should be readily detectable, no high-confidence periodic signal has been found. 
This absence is particularly notable for extremely active repeating FRBs such as FRB 20121102A, FRB 20201124A, FRB 20220912A, and the recently discovered FRB 20240114A, which have provided abundant observational data suitable for such searches \citep{LiD2021,Xu2022,Zhou2022,ZhangYK2023,Zhang2025,WangP2025}. These sources have been extensively monitored and have been detected over $10^3$ bursts each. In particular, FRB 20240114A has been already detected more than $10^4$ bursts \citep{Zhang2025}. Despite the abundance of data, no short period with high statistical significance consistent with a magnetar spin period has been detected in any of these cases \citep{LiD2021,Niu2022,ZhangYK2023,Zhou2025,WangP2025}. The absence of the detectable magnetar spin periods of these sources poses a significant challenge to the widely discussed scenario in which FRBs originate from magnetars. Since more data samples seem to be unhelpful for short-period searches, methods that involve periodic searching through reasonable grouping have started to be used in the search of the short period. Considering that the period of the TOA sequence might depend on the energy of the bursts, \cite{WangP2025} divided 555 bursts of FRB 20121102A observed by FAST into three groups according to their energy to search the short period. As a result, no short periods with high confidence were found during the search. \cite{Jahns2023} presented 849 bursts from FRB 20121102A detected with the 305-m Arecibo Telescope and they performed the phase-folding method and LSP to search the short period of two days with the highest number of detected bursts. 

Notably, \cite{Du2025} reported a $1.7$ s candidate period for FRB 20201124A on two observing days containing 25 and 28 bursts. They proposed that an underlying rotational period may become detectable only when the active emission region is temporarily localized. This interpretation provides a testable statistical hypothesis: even if the signal is weak in an individual short burst sequence, a genuine period should be recovered preferentially across many short subsets relative to unrelated trial periods and to an appropriate aperiodic null process. In addition, \cite{Liu2025} searched for possible rotational signatures by fitting the polarization position-angle (PA) evolution of individual FRB bursts with the rotating vector model (RVM). Among 1727 bursts from three active repeating FRBs, they identified 46 bursts whose PA swings could be well described by the RVM. However, the inferred viewing geometries and rotation periods varied substantially from burst to burst and were inconsistent with a single stable configuration. This inconsistency suggests that the magnetosphere of the FRB central engine may be continuously distorted and dynamically reconfigured during burst activity, such that different bursts may originate from different magnetic-field geometries or emission regions. 

%Motivated by this interpretation, we perform periodicity searches for four highly active repeating FRBs using a sliding-window approach. By selecting temporally adjacent bursts within limited time intervals, this method is designed to identify transient periodic signals that may become detectable when the bursts originate from a relatively localized magnetospheric region. We use the resulting period candidates and their statistical properties to test whether such localized emission episodes can consistently reveal an underlying rotational periodicity, or whether the apparent signals can instead be explained by the intrinsic temporal clustering of FRB bursts.}%However, the statistical significance of this period for the two days remains under debate \citep{Gazith2025}. 

In this study, we search four extremely active repeating FRBs observed with FAST using a sliding-window analysis of temporally adjacent bursts. Rather than treating the resulting candidate peaks as periodicity detections, we use their repeated recovery to test whether a unique underlying period is preferentially selected, as required by the above interpretation. We then compare the observed candidate distributions with homogeneous-Poisson simulations, nonparametric waiting-time randomizations, and epidemic-type aftershock sequence (ETAS) simulations to determine whether the apparent periodicities can instead be generated by the burst waiting-time statistics. This Letter is organized as follows. In Section~\ref{secdata}, we introduce the data sample used in the analysis. In Section~\ref{sec2}, we describe the method used to search the period. In Section~\ref{sec3}, we present the result of the periodicity search. Finally, the discussion and conclusions are presented in Section~\ref{sec4}. 
%The central aim is therefore to test the validity of the significance calibration and the periodic interpretation, rather than to introduce the sliding-window method itself as a new detection technique.

\section{Data sample}\label{secdata}
In this study, we use the data from four active repeating FRB sources: FRB 20121102A, FRB 20201124A, FRB 20220912A, and FRB 20240114A. To avoid potential biases introduced by different observing frequency bands in the periodic activity window such as the chromatic activity window of FRB 20180916B and FRB 20121102A \citep{Pastor-Marazuela2021,Espinoza-Dupouy2025}, we used only observations conducted with the FAST telescope \citep{LiD2021,Xu2022,Zhou2022,ZhangYK2023,Zhang2025,WangP2025}. Specifically, FRB 20121102A contains a total of 2207 bursts over 65 observing days, FRB 20201124A contains 2744 bursts over 49 observing days, FRB 20220912A contains 1076 bursts over 17 observing days, and FRB 20240114A contains 11553 bursts over 59 observing days.The burst catalogs used in this work were obtained from publicly available data releases of previous observational studies. The original data sources and access links are provided in the corresponding references.

\section{Methods}\label{sec2}
%However, the sliding-window samples considered in this work contain only a small number of discrete burst arrival times rather than densely sampled time-series measurements. 
%Therefore, methods specifically developed for sparse event data, such as the Rayleigh $Z_m^2$ periodogram and the H-test, are more appropriate than binned approaches that rely on Gaussian statistics. 
%These unbinned methods directly utilize the phases of individual bursts and avoid additional assumptions related to phase binning.
Various methods can be employed to search for periodic signals in FRB burst arrival times \citep{Chime/FrbCollaboration2020,Niu2022}. However, the sliding-window approach adopted in this work often results in burst samples containing only a limited number of events. In such cases, the number of bursts in individual phase bins ($N_j$) can become small, making the Gaussian approximation underlying the conventional phase-folding $\chi^2$ test potentially unreliable. In addition, methods such as the LSP, which are primarily developed for regularly sampled or continuous time-series data, may not be optimal for sparse event data such as FRB burst arrival times.

To overcome these limitations, we mainly adopt the Rayleigh $Z_m^2$ periodogram and the H-test \citep{Buccheri1983,1989A&A...221..180D}. These method are unbinned and directly utilize the phases of individual bursts, avoiding the dependence on phase-bin selection and the associated low-count statistics \citep{Leahy1983}. Moreover, by incorporating multiple harmonics, the H-test remains sensitive to both sinusoidal and non-sinusoidal phase distributions without requiring the pulse profile to be specified a priori. Therefore, these methods provide a more appropriate statistical framework for searching for short-timescale periodicity in the small-number burst samples produced by our sliding-window analysis.

\subsection{The Rayleigh $Z_m^2$ and H-test method for periodicity analysis}\label{sec_ZF}
%We also applied the Rayleigh $Z_m^2$ test combined with the H-test as an independent cross-check of the phase-folding results. 
%Various methods can be employed to search for periodic signals in FRB burst arrival times \citep{Chime/FrbCollaboration2020,Niu2022}.
Unlike the phase-folding $\chi^2$ test, the $Z_m^2$ statistic is unbinned and is calculated directly from the phases of individual bursts \citep{Buccheri1983,Protheroe1987}. It therefore avoids dependence on the choice of phase bins and alleviates the problem of low event occupancy in individual bins, which is particularly important for the relatively small burst samples considered in our sliding-window analysis.
The method begins by folding the time series data or the TOA data of FRBs into phase space based on a trial period (p) as follows:
\begin{equation}\label{Eq_fold}
    \phi_i=\frac{(t_i-t_0)\ \mod p}{p},
\end{equation}
where $t_i$ represents the TOA of the $i$-th burst, $\phi_i$ is the phase of this burst and $t_0$ is the zero phase point, usually set as the TOA of the first burst. For a trial period $P$, the $Z_m^2$ statistic is then defined as
\begin{equation}
Z_m^2(P)=\frac{2}{N}
\sum_{k=1}^{m}
\left[
\left(\sum_{i=1}^{N}\cos k\phi_i\right)^2
+
\left(\sum_{i=1}^{N}\sin k\phi_i\right)^2
\right],
\end{equation}
where $N$ is the total number of bursts and $m$ is the number of harmonics included in the test. For $m=1$, this reduces to the Rayleigh $Z_1^2$ test, which is most sensitive to approximately sinusoidal phase modulation. Increasing $m$ incorporates higher harmonics and improves the sensitivity to nonsinusoidal or multi-peaked phase distributions.

Under the null hypothesis that the burst phases are independently and uniformly distributed, $Z_m^2$ asymptotically follows a chi-squared distribution with $2m$ degrees of freedom,
\begin{equation}
Z_m^2 \sim \chi_{2m}^{2}.
\end{equation}
For $m=1$, the corresponding single-trial false-alarm probability is
\begin{equation}
p_{\rm single}=\exp\left(-\frac{Z_1^2}{2}\right).
\end{equation}
A large value of $Z_m^2$ therefore indicates that the burst phases are more strongly clustered than expected for a uniform phase distribution at the corresponding trial period.

Because the optimal number of harmonics is generally unknown a priori, we further adopted the H-test, which combines the $Z_m^2$ statistics calculated with different values of $m$ while penalizing the inclusion of additional harmonics \citep{1989A&A...221..180D,Swanepoel1996}. In this work, we considered $m=1$--5 and defined
\begin{equation}
H(P)=\max_{1\leq m\leq5}
\left[Z_m^2(P)-4m+4\right].
\end{equation}
The harmonic number that maximizes the above expression is selected automatically for each trial period. This procedure allows the search to remain sensitive to both nearly sinusoidal signals, which are generally favored by small $m$, and more complicated phase profiles containing significant harmonic structure, without fixing the pulse morphology in advance. We restrict the search to $m\leq5$ because of the relatively small number of bursts contained in individual sliding windows; higher-order harmonics are poorly constrained in such sparse samples and can increase sensitivity to statistical fluctuations.

\subsection{Periodicity Search with Burst Windows}\label{sec_sw}
To test whether an underlying period can be preferentially recovered during short activity intervals, we perform periodicity searches on successive groups of $N$ temporally adjacent bursts within each observing day. We set $N=20$--$30$, providing enough events for the periodicity statistics while retaining samples comparable in size to the 25 and 28 bursts used by \cite{Du2025}. We search periods from $0.01$ s to $100$ s in four intervals: $0.01$--$0.1$ s, $0.1$--$1$ s, $1$--$10$ s, and $10$--$100$ s, using fixed steps of $1\times10^{-5}$ s, $1\times10^{-4}$ s, $1\times10^{-3}$ s, and $1\times10^{-2}$ s, respectively. Note that, due to the limited duration of some burst groups, the upper bound of the longest-period interval is reduced when necessary so that a candidate period spans at least three complete cycles.

\subsection{Poisson-based significance calibration}\label{sec_sig}

We estimated the statistical significance of each periodicity candidate using Monte Carlo simulations under the null hypothesis that the burst arrival times follow a homogeneous Poisson process. For each observed burst subset, we generated synthetic burst sequences containing the same number of bursts as the real data and spanning the same total duration. The synthetic burst arrival times were randomly drawn within the corresponding observing interval and were then analyzed using exactly the same periodicity-search procedure as that applied to the observed data.

For each simulated realization, we searched over the full trial-period range and recorded the maximum value of the corresponding test statistic, i.e., the maximum $H$ value for the H-test or the maximum reduced $\chi^2$ for the phase-folding method. By retaining the maximum statistic from each realization, the resulting null distribution naturally accounts for the multiple trial periods searched within each burst subset.

After repeating this procedure a large number of times, we constructed the empirical null distribution of the maximum test statistic. The false-alarm probability (FAP) of an observed candidate was then estimated as
\begin{equation}
p_{\rm FAP} =
\frac{N_{\rm sim}(S_{\rm max}\geq S_{\rm obs})}
{N_{\rm sim}},
\end{equation}
where $S_{\rm obs}$ is the test statistic of the observed candidate, $S_{\rm max}$ is the maximum statistic obtained in each simulated realization, $N_{\rm sim}$ is the total number of simulations, and $N_{\rm sim}(S_{\rm max}\geq S_{\rm obs})$ is the number of simulations whose maximum statistic is equal to or larger than the observed value.

The corresponding Gaussian-equivalent significance was obtained by converting the false-alarm probability to an equivalent significance level in units of $\sigma$. In this work, we performed up to $10^{7}$ Monte Carlo realizations for the significance calibration. The resulting empirical thresholds were also used to identify period candidates exceeding a specified significance level, such as $3\sigma$.

To construct a reproducible sample of candidate peaks, we define a candidate as a local maximum exceeding the empirical $3\sigma$ threshold derived from the Poisson calibration. Because an independent Monte Carlo calibration for every overlapping window is computationally impractical, we construct a calibration grid in burst number $N$ and duration $T$.

The duration grid contains 100 representative values spanning the observed sliding-window durations, and $N=20$--$30$ matches the burst numbers used in the search. For every $(N,T)$ combination and period-search interval, $10^7$ Poisson realizations are generated, the maximum search statistic is recorded, and the $99.865\%$ percentile is adopted as the nominal $3\sigma$ candidate-selection threshold.

This calibration accounts for the look-elsewhere effect from scanning trial periods within an individual window, but it is not the global post-trial significance of the complete sliding-window search and it assumes homogeneous Poisson arrivals. We therefore refer to these values as nominal Poisson-calibrated significances. The $3\sigma$ level is used only to select a reproducible candidate sample; its adequacy as a detection significance is tested below using null sequences that preserve the observed waiting-time statistics.

\section{Results}\label{sec3}

\subsection{Candidate period}\label{sec3.1}
\begin{figure}
	\centering
	\includegraphics[width=0.8\columnwidth]{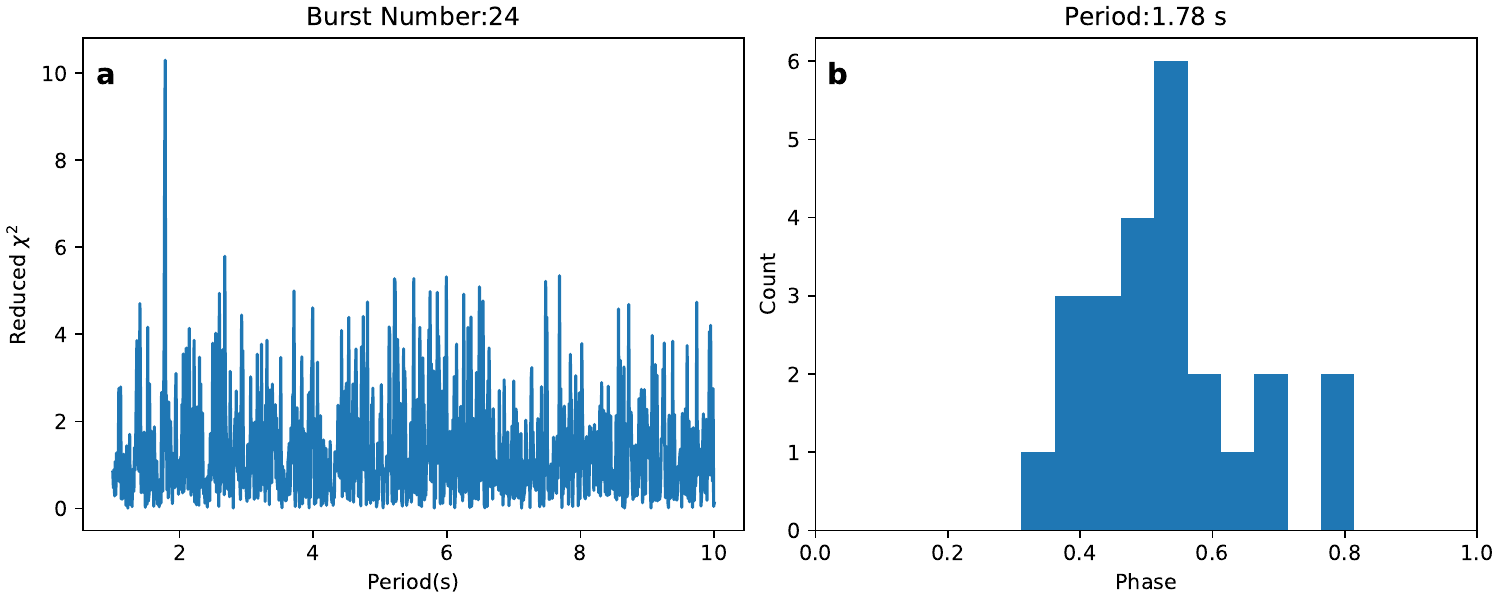}
	\caption{\textbf{The periodogram and folded phase histogram of the $1.78$ s candidate from FRB 20201124A. The quoted $3.62\sigma$ value is its nominal Poisson-calibrated significance, not a global detection significance.}}\label{fig_17_example}
\end{figure}

Our search yields a series of apparent period candidates. For FRB 20201124A, we recover a candidate near $1.78$ s on MJD 59325 from 24 bursts spanning 7029 s (Fig.~\ref{fig_17_example}). Under the homogeneous-Poisson calibration described above, its false-alarm probability corresponds to a nominal Poisson-calibrated significance of $3.62\sigma$, numerically comparable to the $3.9\sigma$ and $3.1\sigma$ values reported by \cite{Du2025}. This value is used here only to characterize the candidate under the same class of Poisson-based calibration, not as a global detection significance.

\begin{figure}
	\centering
	\includegraphics[width=0.8\columnwidth]{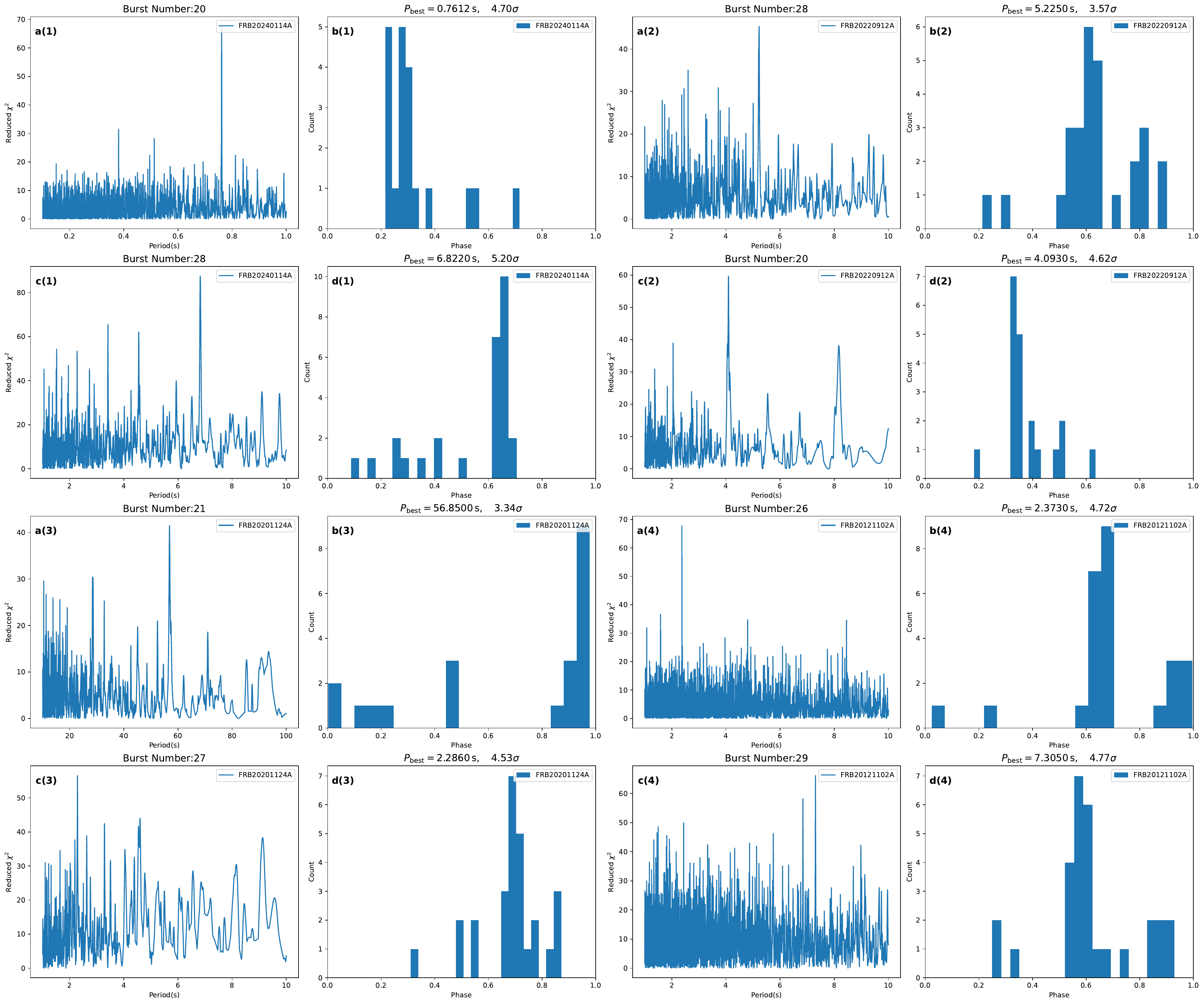}
	\caption{\textbf{Representative periodograms and folded phase histograms of eight apparent period candidates from the four active repeating FRBs. The largest values exceed $5\sigma$ under the nominal Poisson calibration.}}\label{fig_example}
\end{figure}

%The application of this method resulted in the detection of a significant number of periodic candidates with different periods for each repeating FRB source. 
However, the same search produces numerous other candidates with nominal Poisson-calibrated significances above $3\sigma$, spanning periods from $0.01$ s to $100$ s and varying strongly among burst subsets. Eight examples are shown in Fig.~\ref{fig_example}; the largest nominal Poisson-calibrated significance in our search is $5.2\sigma$. The coexistence of many mutually inconsistent candidates shows that a large nominal significance under the Poisson calibration is not sufficient to identify the magnetar spin period.

\begin{figure}
	\centering
	\includegraphics[width=\columnwidth]{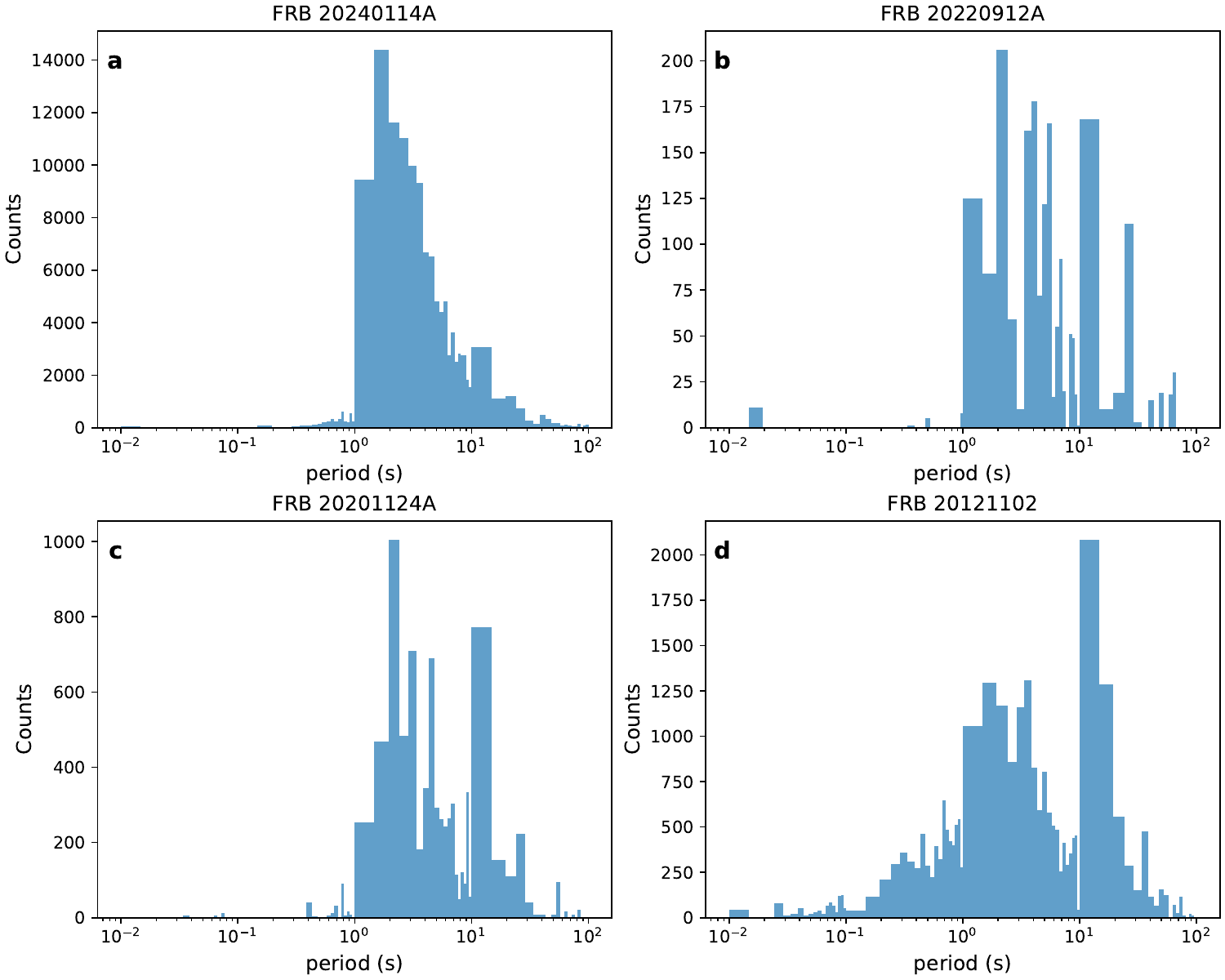}
	\caption{\textbf{Distribution of all candidate peaks selected above the nominal Poisson-calibrated $3\sigma$ threshold for the four active repeating FRBs. The x-axis shows period on a logarithmic scale from $0.01$ s to $100$ s, and the y-axis gives the number of selected candidates in each bin.}}\label{fig_hist_all}
\end{figure}

The existence of many such candidates does not by itself exclude a true periodicity. Under the interpretation motivating \cite{Du2025}, a real period could remain weak in any single short subset but should be recovered preferentially when many subsets are searched. We therefore formulate this as a testable criterion: the true period should have a recurrence probability significantly higher than unrelated trial periods and than that produced by an appropriate aperiodic null process. We use the $>3\sigma$ nominal candidate sample only to construct this recurrence statistic, not as a set of periodicity detections.
%However, if a true period does exist, it should be statistically more likely to be recovered. Based on this consideration, we performed a statistical analysis of the distribution of these period candidates. We then analyzed the distribution of all the period candidates for each repeating FRB source to investigate potential periods or significant features. To further explore possible signals, we relaxed the selection criteria and included candidates with a S/N greater than 5 and a duty cycle less than 0.7.
%As shown in Fig.~\ref{fig_hist1}, there is a clear concentration of period candidates near the period of $1$ s, whereas other period ranges show comparatively fewer detections. 
As shown in Fig.~\ref{fig_hist_all}, the candidate-period distributions of all four active repeating FRB sources show a clear concentration in the range of approximately $1$--$10$ s, although the detailed shapes and amplitudes differ substantially among the sources.

\begin{itemize}
\item FRB 20240114A exhibits the most prominent and sharply concentrated distribution, with the number of period candidates rising rapidly near $\sim1$ s and reaching a maximum of more than $1.4\times10^{4}$ at a period of a few seconds. The distribution then gradually declines toward longer periods, with a weaker extended tail reaching several tens of seconds.

\item FRB 20220912A shows a much more scattered and irregular distribution, with several distinct local peaks between $\sim1$ and $\sim20$ s. The overall number of period candidates is considerably smaller than that of the other three sources, resulting in stronger statistical fluctuations in the histogram. Nevertheless, most candidates are still concentrated within the $1$--$10$ s range.

\item FRB 20201124A also exhibits a pronounced concentration at periods of a few seconds, with the strongest peak occurring near $\sim2$ s and reaching approximately $10^{3}$ counts. Several additional peaks are present at longer periods, particularly around $\sim10$--$20$ s, indicating that the period candidates are not concentrated around a single characteristic value.

\item FRB 20121102A displays a broader and smoother candidate-period distribution than the other sources. The candidates extend from subsecond timescales to several tens of seconds, with substantial concentrations around a few seconds and $\sim10$--$20$ s. The highest peak exceeds $2\times10^{3}$ counts, while the distribution gradually decreases toward both shorter and longer periods.
\end{itemize}

Despite their different detailed structures, the four sources consistently show an enhanced occurrence of period candidates on timescales of a few seconds. As discussed below, however, this common feature does not necessarily indicate an intrinsic periodic timescale, since the recovered distribution may instead be related to the burst waiting-time statistics and the selection effects of the periodicity-search procedure.

%To verify the robustness of this feature, we repeated the analysis across multiple observing days and under varying detection thresholds. The presence of the peak near $1$ s remained consistent, indicating that this phenomenon is broadly reproducible and not an artifact of specific observational conditions.
%\subsection{Periodicity Search with Random Selected Bursts}
\subsection{Homogeneous Poisson simulations}
%A comparison between the results of the simulations and the actual observations is shown in Fig.~\ref{fig_poisson}. As shown in Fig.~\ref{fig_poisson}, the distributions of the simulated and observed results are clearly inconsistent, with significantly fewer detections obtained from the Poisson simulations. This suggests that the burst occurrences of active repeating FRBs do not follow a purely Poissonian process, but instead exhibit a certain degree of temporal correlation \citep{Wang2017,Wang2023,Totani2023}.

To assess the robustness of this finding, we performed Poisson-based simulations as a means of statistical validation. Here, we assume that the occurrence of bursts follows a Poisson process in time, meaning that the bursts are completely independent and randomly distributed. As a first control, we generate homogeneous-Poisson burst sequences using exactly the observed number of bursts and observing duration for each observing day, and analyze them with the identical sliding-window and period-search procedures. Because the simulated and observed data contain the same number of bursts and are partitioned according to the same windowing rule, they contain exactly the same number of searched burst windows. In addition, each window is searched over the same trial-period grid using the same detection statistic and candidate-selection criterion. Therefore, the overall search multiplicity associated with the number of burst windows and the number of trial periods per window is held fixed between the observed and Poisson samples.

This makes the Poisson simulations a useful benchmark for estimating the false candidates generated purely by the large number of trials in our search. In other words, any false-alarm contribution arising simply because many overlapping windows and many trial periods are examined is present in both the real and simulated samples to the same extent. If the large number of candidates in the observations were caused primarily by this search multiplicity, the homogeneous-Poisson simulations should produce a comparable number and distribution of candidates.

\begin{figure}
	\centering
	\includegraphics[width=0.8\columnwidth]{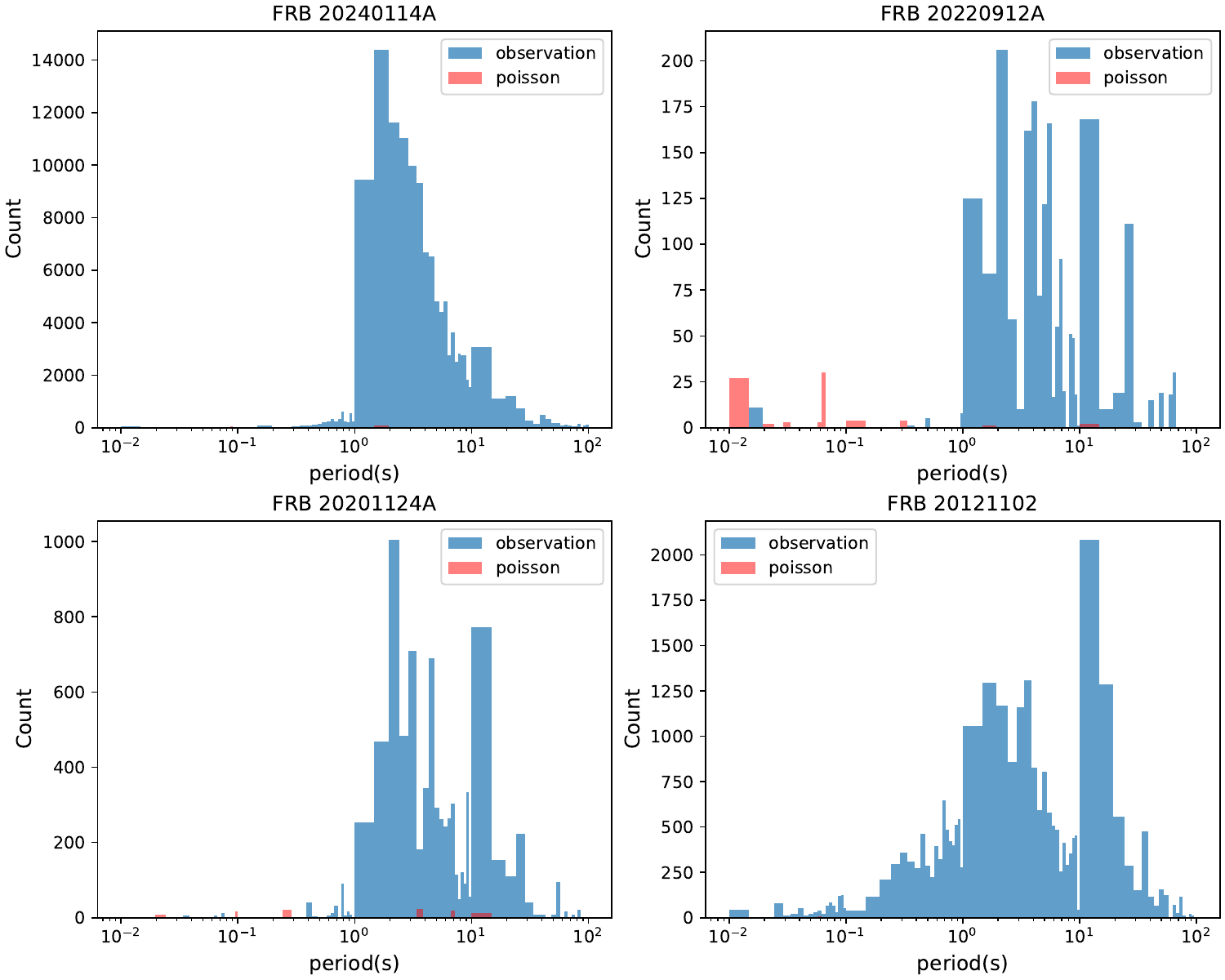}
	\caption{\textbf{Comparison of the candidate-period distributions from the real data and homogeneous-Poisson simulations. Each simulated observing day contains exactly the same number of bursts and duration as the corresponding real day and is processed with the same sliding-window procedure; consequently, the number of searched windows (window trials) is identical in the two cases. The blue histogram shows the real data and the red histogram the Poisson simulations.}}\label{fig_poisson}
\end{figure}

As shown in Fig.~\ref{fig_poisson}, however, the Poisson simulations produce substantially fewer candidates and a markedly different candidate-period distribution, despite having exactly the same number of window trials and being subjected to the same period-search procedure. Fig.~\ref{fig_poisson} therefore provides a direct baseline for the false-alarm rate expected from trial multiplicity under the homogeneous-Poisson null. The excess of candidates in the real data cannot be attributed simply to the number of trials. Instead, it demonstrates that a homogeneous Poisson process fails to reproduce the relevant inter-arrival statistics of the observed bursts, although this discrepancy alone does not imply serial correlation between successive waiting times.

\subsection{Waiting Time Randomization Test}

We next perform a nonparametric waiting-time randomization test that does not assume a specific physical point-process model. For each FRB source, we performed this randomization separately for the bursts detected on each observing day. Specifically, we preserved the waiting-time distribution within each day but randomly shuffled the order of the waiting times to construct new synthetic burst sequences, thereby avoiding contamination from the much longer inter-day waiting times. This preserves the empirical waiting-time distribution but destroys the original temporal ordering and any coherent periodic phase relation. We then applied the same sliding-window periodicity-search procedure to these randomized sequences. As shown in Fig.~\ref{fig_waiting}, the resulting candidate-period distributions remain broadly consistent with those obtained from the original burst sequences, despite the temporal ordering of the bursts being destroyed. This result indicates that the overall structure of the candidate-period distribution is closely coupled to the waiting-time statistics of the bursts, rather than being primarily determined by a persistent intrinsic periodicity. In other words, similar concentrations of apparently significant period candidates can be reproduced even after removing the original temporal sequence, as long as the waiting-time distribution is retained. This suggests that the non-Poissonian and clustered temporal behavior of repeating FRBs can itself introduce preferred timescales into periodicity searches, potentially producing apparently significant signals that do not correspond to the spin period of the central engine. The implications of this effect for the interpretation and statistical significance of short-period candidates are discussed in the following section. 

Nevertheless, we also repeated the sliding-window periodicity search using the phase-folding method and performed the corresponding random waiting-time tests. We find that the candidate-period distributions obtained from phase-folding are not identical to those obtained from the $Z_m^2$/H-test. This difference is expected because the phase-folding method relies on the distribution of bursts among discrete phase bins, and the relatively small number of bursts in individual sliding windows can lead to significant fluctuations in the bin occupancy $N_j$. Therefore, phase-folding can be more sensitive to the choice of phase bins and small-number statistics. Importantly, despite the differences in the detailed distributions, the comparison between the observed burst sequences and the random waiting-time realizations leads to the same qualitative conclusion. As shown in Fig.~\ref{fig_waiting_pf}, the phase-folding candidate-period distributions of the randomized sequences remain broadly similar to those of the original observations for all four FRB sources. This indicates that the recovered candidate-period distribution is mainly determined by the underlying waiting-time statistics rather than the original temporal ordering of bursts. Therefore, although the absolute candidate-period distributions depend on the adopted periodicity-search statistic, our main conclusion that the apparent short-period signals are closely related to the non-Poissonian temporal clustering of FRB bursts remains unchanged.

\subsection{ETAS simulation}\label{sec3.3}
Having established the above result nonparametrically, we use ETAS as a complementary parametric illustration of a clustered point process rather than as the primary null model \citep{Ogata1988,Ogata1998,Luo2025}. In the rotation-modulated starquake framework of \citet{Luo2025}, the ETAS process describes self-exciting event occurrence: event magnitudes (equivalently, logarithmic energies) follow an approximately exponential Gutenberg--Richter-like distribution, while the rate of triggered events decays as a power law in time after a parent event (an Omori-type law), producing broad non-exponential waiting-time statistics. Rotational modulation and viewing geometry are then applied to determine which events are observed.

\begin{figure}
	\centering
	\includegraphics[width=0.8\columnwidth]{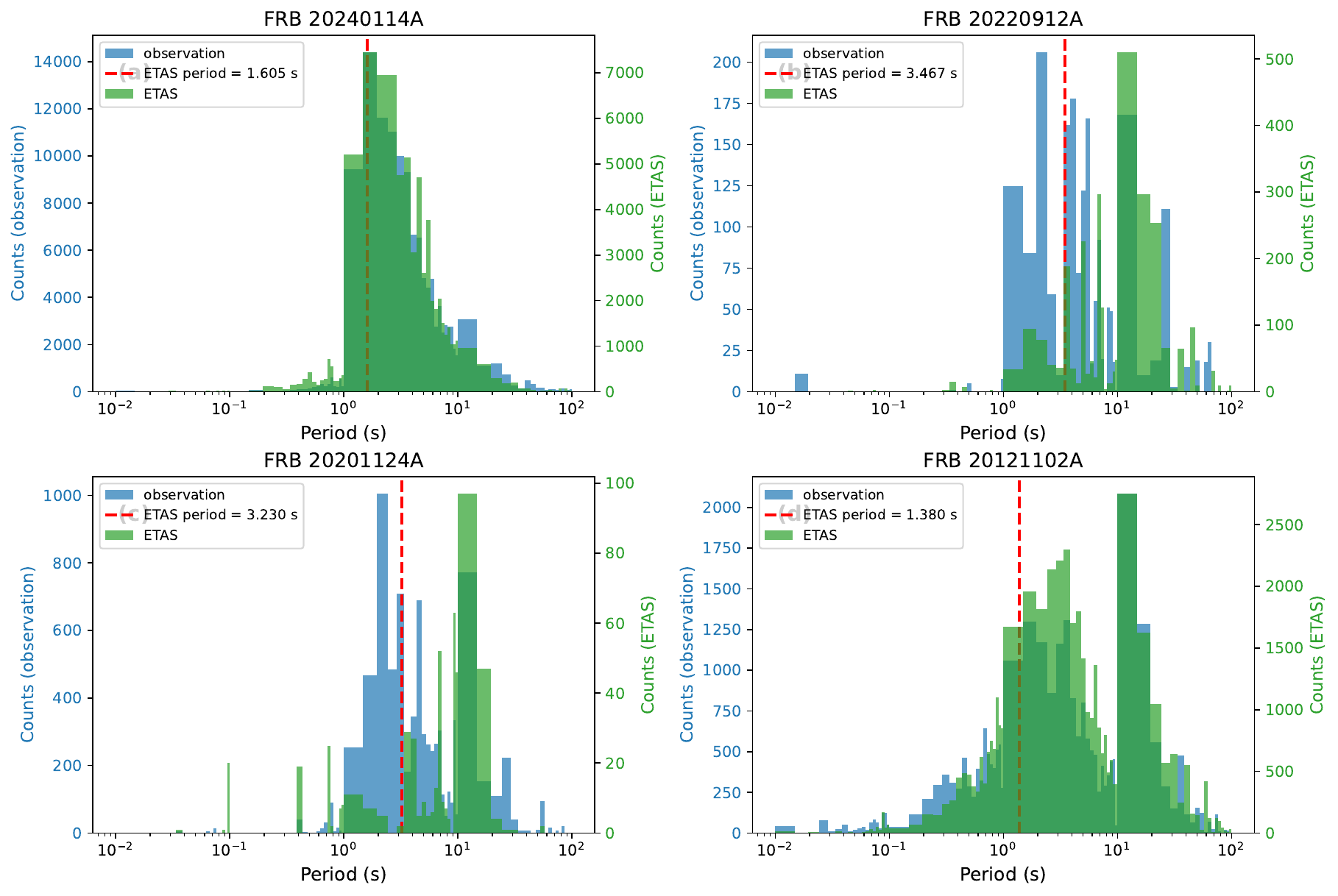}
	\caption{Comparison of period candidate distributions with ETAS simulation and real data. The blue histogram corresponds to results obtained using the real data, while the red histogram represents results from ETAS simulation. The red dashed line indicates the period injected into the ETAS model.}\label{fig_starquake}
\end{figure}

We adopted the same method as \cite{Luo2025} for the ETAS simulations. For FRB 20201124A, FRB 20220912A, and FRB 20121102A, we adopted the same parameters as those used by \cite{Luo2025}. For FRB 20240114A, which was not modeled in \cite{Luo2025}, we performed the MCMC fitting following a similar approach. The best fit parameters are presented in the Table~\ref{table1}. 
After completing the ETAS simulations, we applied the same periodicity-search procedure to the simulated burst sequences and compared the resulting candidate-period distributions with those derived from the observations. 
As shown in Fig.~\ref{fig_starquake}, the blue and green histograms show the candidate-period distributions obtained from the observations and ETAS simulations, respectively. The ETAS simulations broadly reproduce the observed candidate-period distributions for FRB 20240114A and FRB 20121102A, whereas the agreement is substantially poorer for FRB 20220912A and FRB 20201124A. In addition, for FRB 20121102A, FRB 20220912A, and FRB 20201124A, the dominant peaks of the simulated distributions do not coincide well with the injected periods, suggesting that the recovered distributions are not primarily determined by the true periodicity. However, for FRB 20240114A, the ETAS simulation shows good agreement with the injected period.

\begin{figure}
	\centering
	\includegraphics[width=0.8\columnwidth]{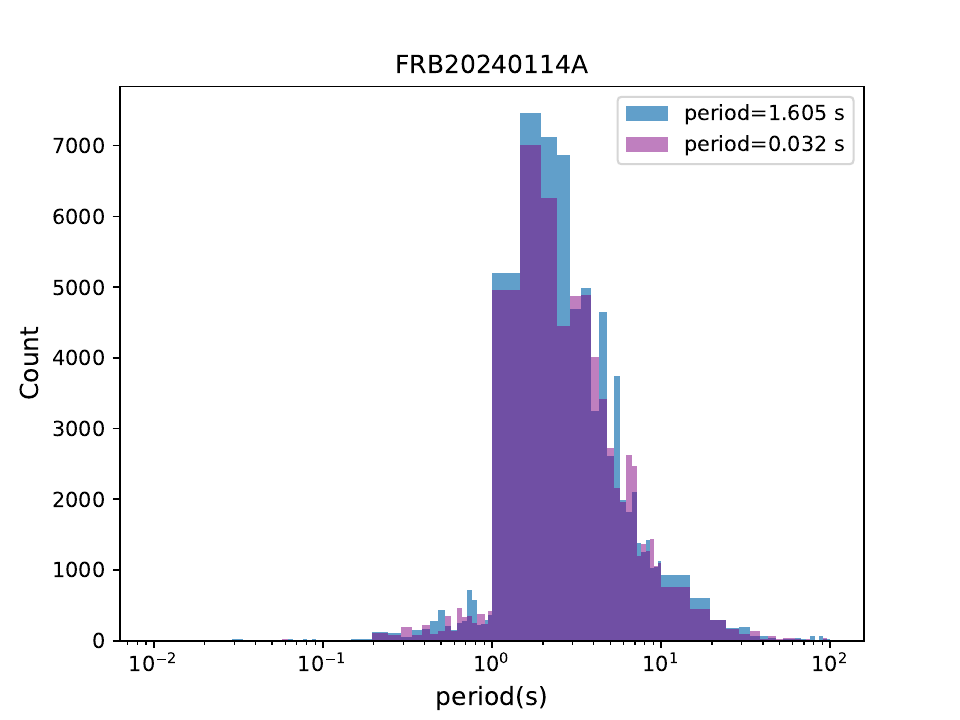}
	\caption{Comparison of period candidate distributions with ETAS simulation and real data. The blue histogram corresponds to results obtained using the real data, while the red histogram represents results from ETAS simulation.}\label{fig_etas_period}
\end{figure}

To further investigate whether the prominent peak identified for FRB 20240114A represents a genuine periodic signal, we performed additional ETAS-based simulations to test whether the recurrence distribution of period candidates can reliably trace an underlying periodicity. We varied only the periodicity parameter in the ETAS model while keeping all other parameters fixed, and repeated the same sliding-window periodicity-search procedure. If the recurrence distribution is indeed sensitive to a genuine periodicity, the recovered candidates should preferentially cluster around the injected period, and the location of this excess should shift accordingly when the injected period is changed. This injection-and-recovery test therefore provides a direct assessment of whether the observed recurrence pattern can be interpreted as evidence for an intrinsic periodic signal.
However, as shown in Fig.~\ref{fig_etas_period}, we find that the recurrence distribution of the recovered period candidates remains essentially unchanged when the injected period is varied. In particular, the prominent features in the distribution do not shift toward the corresponding injected periods, as would be expected if they were tracing a genuine periodic signal. This insensitivity to the injected period indicates that the observed distribution is unlikely to be determined by an intrinsic periodicity. Instead, it is more plausibly produced by the burst waiting-time statistics together with the sampling and sliding-window search procedure.

\section{Discussion}\label{sec4}
The main result of this work is not the identification of additional short periods, but the demonstration that a homogeneous-Poisson calibration does not provide a reliable global significance for clustered FRB event data. The many nominal $>3\sigma$ candidates, including the $1.78$ s candidate in FRB 20201124A, are mutually inconsistent and their recurrence structure can be reproduced by aperiodic surrogate sequences that preserve the observed waiting-time statistics. We therefore find no statistically robust short-period periodicity in the data analyzed here.

This also clarifies why we examine candidates below conventional pulsar or FRB detection thresholds. We do not regard an individual nominal $3\sigma$ peak as evidence for a period. Rather, following the hypothesis motivating \cite{Du2025}, we test whether a true period would be preferentially recovered across many short burst subsets. Such a period should produce a recurrence excess above competing periods and above an appropriate aperiodic null. The injection tests and waiting-time randomizations show that the observed excesses do not satisfy this criterion.

\begin{figure}
	\centering
	\includegraphics[width=0.8\columnwidth]{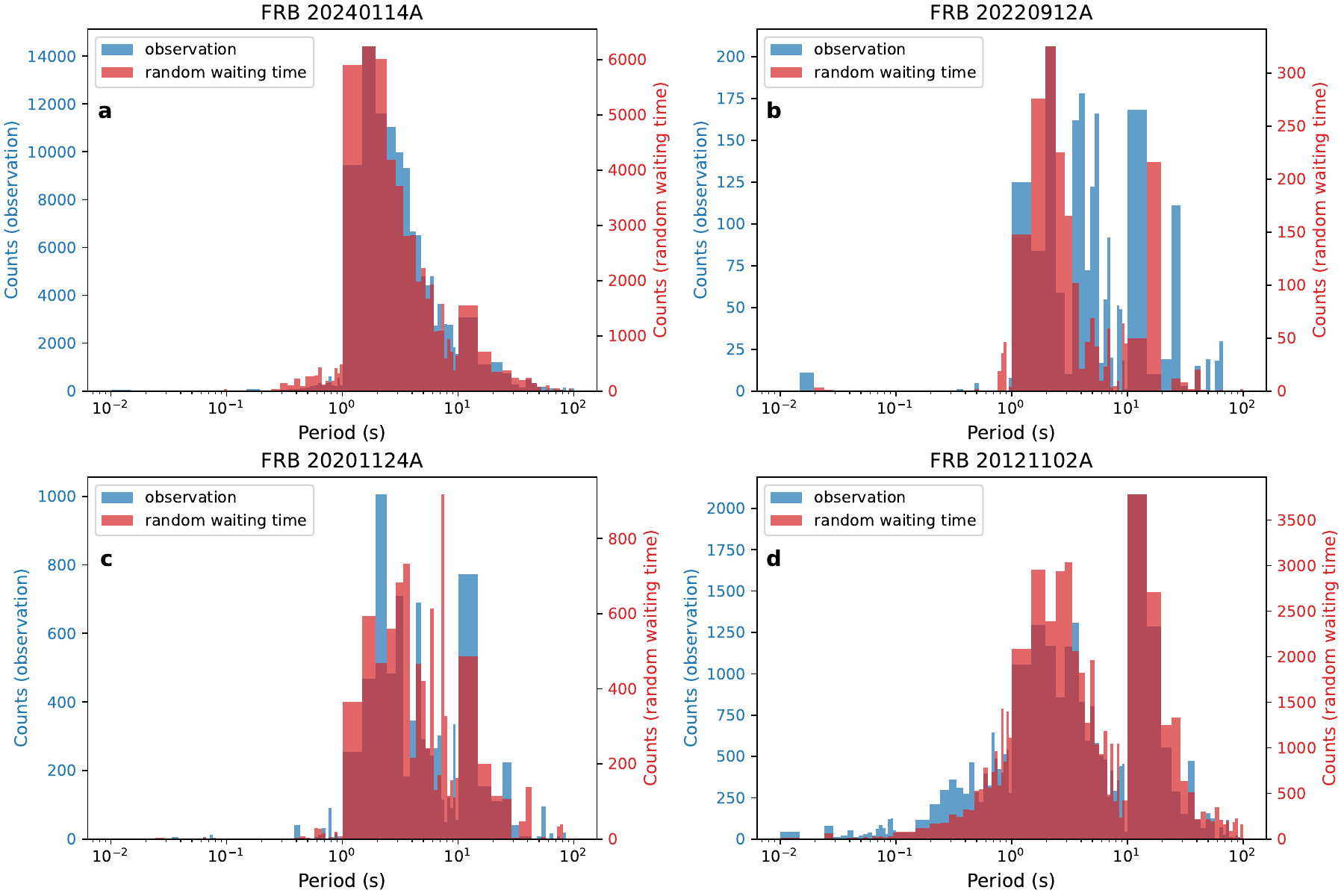}
	\caption{\textbf{Comparison of candidate-period distributions from the real data and the waiting-time-randomized sequences. The empirical waiting times within each observing day are preserved but randomly reordered before applying the same sliding-window search. The blue histogram shows the real data and the red histogram the randomized sequences.}}\label{fig_waiting}
\end{figure}

\begin{figure}
	\centering
	\includegraphics[width=0.8\columnwidth]{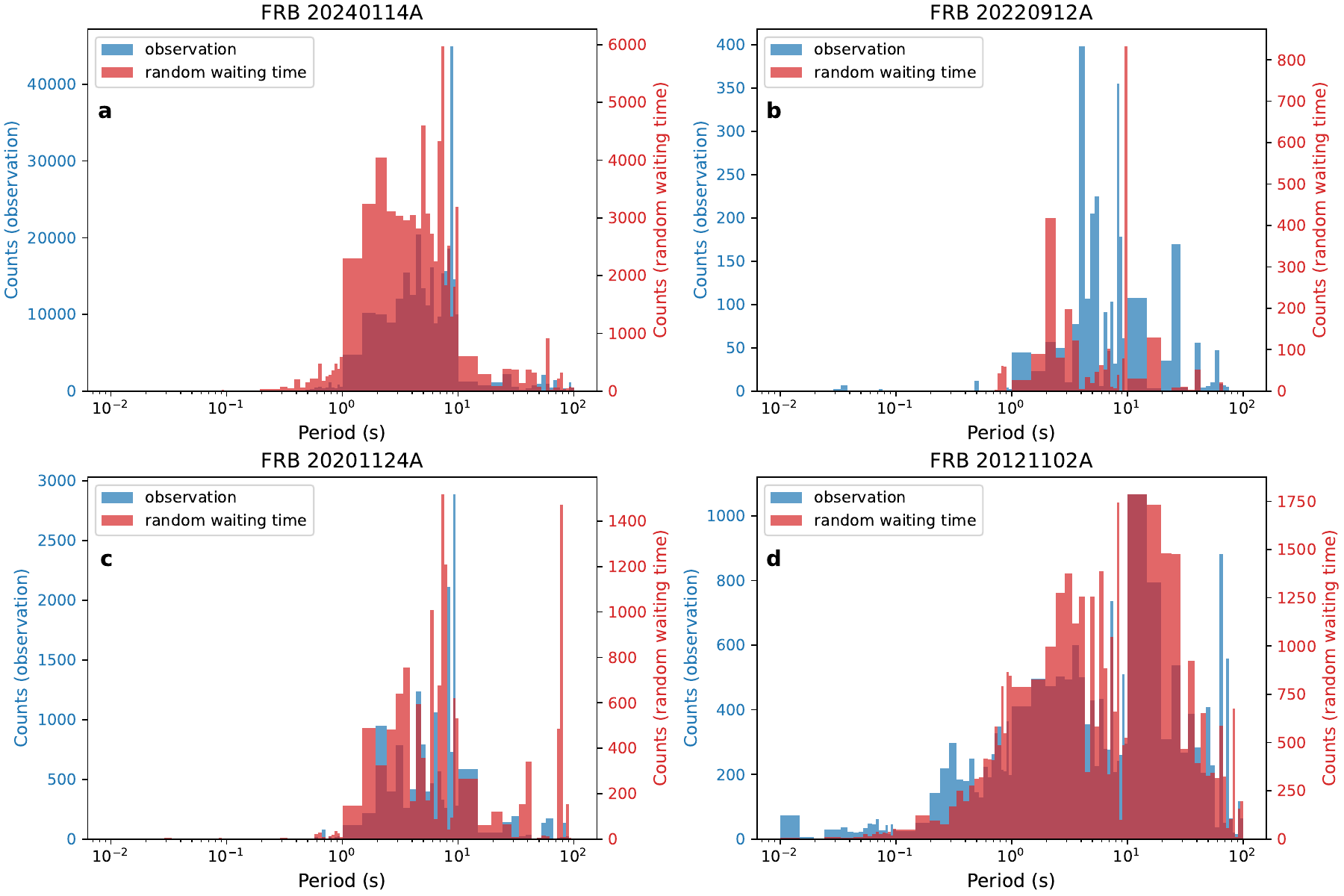}
	\caption{\textbf{Same as Fig.~\ref{fig_waiting}, but using the phase-folding statistic. The broad agreement between the observed and waiting-time-randomized distributions is preserved despite differences in the detailed candidate distribution.}}\label{fig_waiting_pf}
\end{figure}

Figure~\ref{fig_poisson} provides a first control based on homogeneous-Poisson burst sequences, constructed with the same number of bursts, observing durations, and searched windows as the real data. It therefore isolates the level of apparent candidates expected from the search procedure and trial multiplicity under a Poisson null. More importantly, Fig.~\ref{fig_waiting} presents our nonparametric waiting-time randomization test, in which the observed waiting times are retained but their order is shuffled. This preserves the empirical waiting-time distribution while removing the original temporal ordering and any coherent periodic phase information, allowing us to test directly whether the observed candidate structure requires an intrinsic periodic signal.

The contrast between Figs.~\ref{fig_poisson} and \ref{fig_waiting} is particularly informative. Replacing the observations with homogeneous-Poisson arrivals, whose waiting times are exponential, changes both the number and distribution of the candidates even though the number of searched windows is identical. By contrast, permuting the observed waiting times destroys their original ordering but preserves their empirical distribution and reproduces the observed candidate structure much more closely. This nonparametric comparison provides the strongest evidence in our analysis that the apparent candidate-period concentrations are primarily controlled by the waiting-time statistics rather than by a coherent periodic signal. The result is therefore not a generic consequence of randomizing the TOAs; the non-Poissonian waiting-time distribution is the essential ingredient.

%\textbf{The contrast between Figs.~\ref{fig_poisson} and \ref{fig_waiting} is particularly informative. Replacing the observations with homogeneous-Poisson arrivals, whose waiting times are exponential, changes both the number and distribution of the candidates even though the number of searched windows is identical. By contrast, permuting the observed waiting times destroys their original ordering but preserves their empirical distribution and reproduces the observed candidate structure much more closely. The result is therefore not a generic consequence of randomizing the TOAs; the non-Poissonian waiting-time distribution is the essential ingredient.}

The importance of the waiting-time distribution is well established in the broader stochastic-process literature. A homogeneous Poisson process is a special case characterized by independent, exponentially distributed inter-event times, whereas renewal theory allows general inter-arrival distributions and point-process theory further accommodates clustered and history-dependent event sequences \citep{Cox1962,DaleyVereJones2003}. General inter-arrival distributions are also routinely considered in queueing theory and operations research \citep{Asmussen2003}. Consequently, non-exponential waiting-time statistics can alter event-count fluctuations and introduce characteristic stochastic timescales without requiring an underlying periodic clock. In this context, our waiting-time randomization provides a finite-sample, nonparametric renewal-like surrogate: it preserves the empirical set of inter-burst intervals while removing their original temporal ordering.

Closely related model-checking problems arise in other point-process applications. In earthquake statistics, residual analyses based on transforming a fitted conditional-intensity model have long been used to test whether the remaining event sequence is consistent with a Poisson process \citep{Ogata1988}. Similar ideas are widely used in neural spike-train analysis, where the time-rescaling theorem provides a goodness-of-fit test by requiring the rescaled inter-event intervals of a correctly specified point-process model to be independent and exponentially distributed \citep{Brown2002}; conditional-intensity models can additionally incorporate explicit dependence on the previous event history \citep{Truccolo2005}. These approaches emphasize a general point relevant to FRB periodicity searches: the stochastic model used as the null hypothesis for the burst arrival process should be validated before the significance of an apparent periodic signal is interpreted.

To further examine the origin of the apparent candidate-period structures, we modeled the burst sequences using an ETAS framework to reproduce the clustered waiting-time behavior, combined with an RVM geometry that allows an intrinsic rotation period to be injected into the simulated burst sequence. Applying the same sliding-window search, we find that the ETAS simulations reproduce part of the observed candidate-period structure. More importantly, when the injected rotation period is varied, the recovered candidate-period distributions remain largely unchanged rather than shifting with the imposed period. This indicates that the prominent candidate structures are primarily associated with the underlying waiting-time statistics rather than with the injected rotational signal. ETAS therefore provides a model-dependent illustration of how clustered waiting-time behavior can generate apparent periodic signatures, complementing the more general result obtained from the nonparametric waiting-time randomization test.

If an ETAS-like process provides an adequate phenomenological description of repeating FRB activity, it may point to a cascading or self-exciting energy-release mechanism in the magnetar environment. In such a scenario, an initial crustal failure or magnetic-energy release could redistribute stresses and increase the probability of subsequent events, producing temporally clustered bursts analogous to earthquake aftershock sequences. This interpretation is compatible with starquake-driven magnetar activity, but is not unique to crustal failures; correlated magnetospheric reconnection or other time-dependent triggering processes may generate similar statistics. Our results therefore do not establish ETAS as the underlying physical mechanism. Rather, they show that physically plausible clustered burst activity can produce apparent short-period candidates without requiring a persistent rotational modulation. Further tests involving energy-dependent triggering, post-burst rate evolution, and event-history correlations are needed to distinguish between competing physical interpretations.

The localized-emission interpretation proposed by \cite{Du2025} remains a physical motivation for searching short burst subsets, but the present data do not distinguish that scenario from waiting-time-induced false periodicity. In particular, recovering a candidate near $1.7$ s is not sufficient evidence for the neutron-star spin period when many other nominally significant periods are recovered and the same recurrence behavior arises under the waiting-time null.

\section{Conclusions}\label{sec5}

We searched for short-timescale periodicity in four active repeating FRBs with a sliding-window analysis and found numerous apparent candidates, including a $\sim1.78$ s candidate in FRB 20201124A. Their quoted $\sigma$ values are nominal significances calibrated against a homogeneous-Poisson null and are not the global post-trial significance of the complete search. A genuine period should be preferentially recovered relative to other periods and to an appropriate aperiodic null, but this criterion is not satisfied: the candidate concentrations persist when the observed waiting times are randomly reordered while their empirical distribution is preserved. ETAS simulations provide a complementary parametric example of the same general effect. We therefore find no robust evidence for an intrinsic short periodicity in these data. This conclusion does not imply that previously reported or future short-period signals are necessarily spurious. Rather, our results identify an aperiodic alternative explanation that must be rejected before an apparent periodicity can be securely associated with neutron-star rotation. Future FRB spin-period searches should evaluate significance using null models that reproduce the observed waiting-time statistics and the full search procedure before interpreting apparent short-period candidates as evidence for neutron-star rotation.

\section*{Acknowledgements}
This work was supported by the National Natural Science Foundation of China (grant Nos. 12625304, 12494575 and 12393812). J.-W.~Luo is supported by the National Natural Science Foundation of China (grant No.~12403100). Qin Wu is supported by National Natural Science Foundation of China (grant No. 12503050).

    \bibliographystyle{aasjournal}
    \bibliography{refs}

@ARTICLE{Wang2025,
       author = {{Wang}, F.~Y. and {Lan}, H.~T. and {Zhao}, Z.~Y. and {Wu}, Q. and {Feng}, Y. and {Yi}, S.~X. and {Dai}, Z.~G. and {Cheng}, K.~S.},
        title = "{A Unified Model of Active Repeating Fast Radio Bursts Associated with Persistent Radio Sources}",
      journal = {\apj},
         year = 2026,
        month = oct,
       volume = {1009},
       number = {2},
          eid = {211},
        pages = {211},
          doi = {10.3847/1538-4357/ae9d5a},
archivePrefix = {arXiv},
       eprint = {2512.07140},
 primaryClass = {astro-ph.HE},
       adsurl = {https://ui.adsabs.harvard.edu/abs/2026ApJ..1009..211W}
}

@ARTICLE{Wang2026,
       author = {{Wang}, Fayin and {Jia}, Xuandong and {Gao}, Daohong and {Dai}, Zigao},
        title = "{Fast Radio Burst Cosmology: Hubble Tension and Dark Energy}",
      journal = {Research in Astronomy and Astrophysics},
         year = 2026,
        month = aug,
       volume = {26},
       number = {8},
          eid = {084010},
        pages = {084010},
          doi = {10.1088/1674-4527/ae8428},
archivePrefix = {arXiv},
       eprint = {2606.22390},
 primaryClass = {astro-ph.CO},
       adsurl = {https://ui.adsabs.harvard.edu/abs/2026RAA....26h4010W}
}

@ARTICLE{Rajwade2023,
       author = {{Rajwade}, K.~M. and {van den Eijnden}, J.},
        title = "{Expectations for fast radio bursts in neutron star-massive star binaries}",
      journal = {\aap},
         year = 2023,
        month = may,
       volume = {673},
          eid = {A136},
        pages = {A136},
          doi = {10.1051/0004-6361/202245468},
archivePrefix = {arXiv},
       eprint = {2303.16770},
 primaryClass = {astro-ph.HE},
       adsurl = {https://ui.adsabs.harvard.edu/abs/2023A&A...673A.136R}
}

@ARTICLE{Zhao2023,
       author = {{Zhao}, Z.~Y. and {Zhang}, G.~Q. and {Wang}, F.~Y. and {Dai}, Z.~G.},
        title = "{Rotation Measure Variations and Reversals of Repeating FRBs in Massive Binary Systems}",
      journal = {\apj},
         year = 2023,
        month = jan,
       volume = {942},
       number = {2},
          eid = {102},
        pages = {102},
          doi = {10.3847/1538-4357/aca66b},
archivePrefix = {arXiv},
       eprint = {2209.05242},
 primaryClass = {astro-ph.HE},
       adsurl = {https://ui.adsabs.harvard.edu/abs/2023ApJ...942..102Z}
}

@ARTICLE{ZhangB2025,
       author = {{Zhang}, Bing and {Hu}, Rui-Chong},
        title = "{Magnetars in Binaries as the Engine of Actively Repeating Fast Radio Bursts}",
      journal = {arXiv e-prints},
         year = 2025,
        month = aug,
          eid = {arXiv:2508.12119},
        pages = {arXiv:2508.12119},
          doi = {10.48550/arXiv.2508.12119},
archivePrefix = {arXiv},
       eprint = {2508.12119},
 primaryClass = {astro-ph.HE},
       adsurl = {https://ui.adsabs.harvard.edu/abs/2025arXiv250812119Z}
}

@ARTICLE{Gu2020,
       author = {{Gu}, Wei-Min and {Yi}, Tuan and {Liu}, Tong},
        title = "{A neutron star-white dwarf binary model for periodically active fast radio burst sources}",
      journal = {\mnras},
         year = 2020,
        month = sep,
       volume = {497},
       number = {2},
        pages = {1543-1546},
          doi = {10.1093/mnras/staa1914},
archivePrefix = {arXiv},
       eprint = {2002.10478},
 primaryClass = {astro-ph.HE},
       adsurl = {https://ui.adsabs.harvard.edu/abs/2020MNRAS.497.1543G}
}

@ARTICLE{Metzger2017,
       author = {{Metzger}, Brian D. and {Berger}, Edo and {Margalit}, Ben},
        title = "{Millisecond Magnetar Birth Connects FRB 121102 to Superluminous Supernovae and Long-duration Gamma-Ray Bursts}",
      journal = {\apj},
         year = 2017,
        month = may,
       volume = {841},
       number = {1},
          eid = {14},
        pages = {14},
          doi = {10.3847/1538-4357/aa633d},
archivePrefix = {arXiv},
       eprint = {1701.02370},
 primaryClass = {astro-ph.HE},
       adsurl = {https://ui.adsabs.harvard.edu/abs/2017ApJ...841...14M}
}

@ARTICLE{Cordes2019,
       author = {{Cordes}, James M. and {Chatterjee}, Shami},
        title = "{Fast Radio Bursts: An Extragalactic Enigma}",
      journal = {\araa},
         year = 2019,
        month = aug,
       volume = {57},
        pages = {417-465},
          doi = {10.1146/annurev-astro-091918-104501},
archivePrefix = {arXiv},
       eprint = {1906.05878},
 primaryClass = {astro-ph.HE},
       adsurl = {https://ui.adsabs.harvard.edu/abs/2019ARA&A..57..417C}
}

@ARTICLE{Zhang2023,
       author = {{Zhang}, Bing},
        title = "{The physics of fast radio bursts}",
      journal = {Reviews of Modern Physics},
         year = 2023,
        month = jul,
       volume = {95},
       number = {3},
          eid = {035005},
        pages = {035005},
          doi = {10.1103/RevModPhys.95.035005},
archivePrefix = {arXiv},
       eprint = {2212.03972},
 primaryClass = {astro-ph.HE},
       adsurl = {https://ui.adsabs.harvard.edu/abs/2023RvMP...95c5005Z}
}

@ARTICLE{Lorimer2007,
       author = {{Lorimer}, D.~R. and {Bailes}, M. and {McLaughlin}, M.~A. and {Narkevic}, D.~J. and {Crawford}, F.},
        title = "{A Bright Millisecond Radio Burst of Extragalactic Origin}",
      journal = {Science},
         year = 2007,
        month = nov,
       volume = {318},
       number = {5851},
        pages = {777},
          doi = {10.1126/science.1147532},
archivePrefix = {arXiv},
       eprint = {0709.4301},
 primaryClass = {astro-ph},
       adsurl = {https://ui.adsabs.harvard.edu/abs/2007Sci...318..777L}
}

@ARTICLE{CHIME/FRBCollaboration2020b,
       author = {{CHIME/FRB Collaboration} and {Andersen}, B.~C. and {Bandura}, K.~M. and {Bhardwaj}, M. and {Bij}, A. and {Boyce}, M.~M. and {Boyle}, P.~J. and {Brar}, C. and {Cassanelli}, T. and {Chawla}, P. and {Chen}, T. and {Cliche}, J. -F. and {Cook}, A. and {Cubranic}, D. and {Curtin}, A.~P. and {Denman}, N.~T. and {Dobbs}, M. and {Dong}, F.~Q. and {Fandino}, M. and {Fonseca}, E. and {Gaensler}, B.~M. and {Giri}, U. and {Good}, D.~C. and {Halpern}, M. and {Hill}, A.~S. and {Hinshaw}, G.~F. and {H{\"o}fer}, C. and {Josephy}, A. and {Kania}, J.~W. and {Kaspi}, V.~M. and {Landecker}, T.~L. and {Leung}, C. and {Li}, D.~Z. and {Lin}, H. -H. and {Masui}, K.~W. and {McKinven}, R. and {Mena-Parra}, J. and {Merryfield}, M. and {Meyers}, B.~W. and {Michilli}, D. and {Milutinovic}, N. and {Mirhosseini}, A. and {M{\"u}nchmeyer}, M. and {Naidu}, A. and {Newburgh}, L.~B. and {Ng}, C. and {Patel}, C. and {Pen}, U. -L. and {Pinsonneault-Marotte}, T. and {Pleunis}, Z. and {Quine}, B.~M. and {Rafiei-Ravandi}, M. and {Rahman}, M. and {Ransom}, S.~M. and {Renard}, A. and {Sanghavi}, P. and {Scholz}, P. and {Shaw}, J.~R. and {Shin}, K. and {Siegel}, S.~R. and {Singh}, S. and {Smegal}, R.~J. and {Smith}, K.~M. and {Stairs}, I.~H. and {Tan}, C.~M. and {Tendulkar}, S.~P. and {Tretyakov}, I. and {Vanderlinde}, K. and {Wang}, H. and {Wulf}, D. and {Zwaniga}, A.~V.},
        title = "{A bright millisecond-duration radio burst from a Galactic magnetar}",
      journal = {\nat},
         year = 2020,
        month = nov,
       volume = {587},
       number = {7832},
        pages = {54-58},
          doi = {10.1038/s41586-020-2863-y},
archivePrefix = {arXiv},
       eprint = {2005.10324},
 primaryClass = {astro-ph.HE},
       adsurl = {https://ui.adsabs.harvard.edu/abs/2020Natur.587...54C}
}

@ARTICLE{Bochenek2020,
       author = {{Bochenek}, C.~D. and {Ravi}, V. and {Belov}, K.~V. and {Hallinan}, G. and {Kocz}, J. and {Kulkarni}, S.~R. and {McKenna}, D.~L.},
        title = "{A fast radio burst associated with a Galactic magnetar}",
      journal = {\nat},
         year = 2020,
        month = nov,
       volume = {587},
       number = {7832},
        pages = {59-62},
          doi = {10.1038/s41586-020-2872-x},
archivePrefix = {arXiv},
       eprint = {2005.10828},
 primaryClass = {astro-ph.HE},
       adsurl = {https://ui.adsabs.harvard.edu/abs/2020Natur.587...59B}
}

@ARTICLE{CHIME/FRBCollaboration2019,
       author = {{CHIME/FRB Collaboration} and {Amiri}, M. and {Bandura}, K. and {Bhardwaj}, M. and {Boubel}, P. and {Boyce}, M.~M. and {Boyle}, P.~J. and {Brar}, C. and {Burhanpurkar}, M. and {Chawla}, P. and {Cliche}, J.~F. and {Cubranic}, D. and {Deng}, M. and {Denman}, N. and {Dobbs}, M. and {Fandino}, M. and {Fonseca}, E. and {Gaensler}, B.~M. and {Gilbert}, A.~J. and {Giri}, U. and {Good}, D.~C. and {Halpern}, M. and {Hanna}, D and {Hill}, A.~S. and {Hinshaw}, G. and {H{\"o}fer}, C. and {Josephy}, A. and {Kaspi}, V.~M. and {Landecker}, T.~L. and {Lang}, D.~A. and {Masui}, K.~W. and {Mckinven}, R. and {Mena-Parra}, J. and {Merryfield}, M. and {Milutinovic}, N. and {Moatti}, C. and {Naidu}, A. and {Newburgh}, L.~B. and {Ng}, C. and {Patel}, C. and {Pen}, U. and {Pinsonneault-Marotte}, T. and {Pleunis}, Z. and {Rafiei-Ravandi}, M. and {Ransom}, S.~M. and {Renard}, A. and {Scholz}, P. and {Shaw}, J.~R. and {Siegel}, S.~R. and {Smith}, K.~M. and {Stairs}, I.~H. and {Tendulkar}, S.~P. and {Tretyakov}, I. and {Vanderlinde}, K. and {Yadav}, P.},
        title = "{Observations of fast radio bursts at frequencies down to 400 megahertz}",
      journal = {\nat},
         year = 2019,
        month = jan,
       volume = {566},
       number = {7743},
        pages = {230-234},
          doi = {10.1038/s41586-018-0867-7},
archivePrefix = {arXiv},
       eprint = {1901.04524},
 primaryClass = {astro-ph.HE},
       adsurl = {https://ui.adsabs.harvard.edu/abs/2019Natur.566..230C}
}

@ARTICLE{Chime/FrbCollaboration2023,
       author = {{CHIME/FRB Collaboration} and {Andersen}, Bridget C. and {Bandura}, Kevin and {Bhardwaj}, Mohit and {Boyle}, P.~J. and {Brar}, Charanjot and {Cassanelli}, Tomas and {Chatterjee}, S. and {Chawla}, Pragya and {Cook}, Amanda M. and {Curtin}, Alice P. and {Dobbs}, Matt and {Dong}, Fengqiu Adam and {Faber}, Jakob T. and {Fandino}, Mateus and {Fonseca}, Emmanuel and {Gaensler}, B.~M. and {Giri}, Utkarsh and {Herrera-Martin}, Antonio and {Hill}, Alex S. and {Ibik}, Adaeze and {Josephy}, Alexander and {Kaczmarek}, Jane F. and {Kader}, Zarif and {Kaspi}, Victoria and {Landecker}, T.~L. and {Lanman}, Adam E. and {Lazda}, Mattias and {Leung}, Calvin and {Lin}, Hsiu-Hsien and {Masui}, Kiyoshi W. and {McKinven}, Ryan and {Mena-Parra}, Juan and {Meyers}, Bradley W. and {Michilli}, D. and {Ng}, Cherry and {Pandhi}, Ayush and {Pearlman}, Aaron B. and {Pen}, Ue-Li and {Petroff}, Emily and {Pleunis}, Ziggy and {Rafiei-Ravandi}, Masoud and {Rahman}, Mubdi and {Ransom}, Scott M. and {Renard}, Andre and {Sand}, Ketan R. and {Sanghavi}, Pranav and {Scholz}, Paul and {Shah}, Vishwangi and {Shin}, Kaitlyn and {Siegel}, Seth and {Smith}, Kendrick and {Stairs}, Ingrid and {Su}, Jianing and {Tendulkar}, Shriharsh P. and {Vanderlinde}, Keith and {Wang}, Haochen and {Wulf}, Dallas and {Zwaniga}, Andrew},
        title = "{CHIME/FRB Discovery of 25 Repeating Fast Radio Burst Sources}",
      journal = {\apj},
         year = 2023,
        month = apr,
       volume = {947},
       number = {2},
          eid = {83},
        pages = {83},
          doi = {10.3847/1538-4357/acc6c1},
archivePrefix = {arXiv},
       eprint = {2301.08762},
 primaryClass = {astro-ph.HE},
       adsurl = {https://ui.adsabs.harvard.edu/abs/2023ApJ...947...83C}
}

@ARTICLE{Niu2022,
       author = {{Niu}, Jia-Rui and {Zhu}, Wei-Wei and {Zhang}, Bing and {Yuan}, Mao and {Zhou}, De-Jiang and {Zhang}, Yong-Kun and {Jiang}, Jin-Chen and {Han}, J.~L. and {Li}, Di and {Lee}, Ke-Jia and {Wang}, Pei and {Feng}, Yi and {Li}, Dong-Zi and {Luo}, Rui and {Wang}, Fa-Yin and {Dai}, Zi-Gao and {Miao}, Chen-Chen and {Niu}, Chen-Hui and {Xu}, Heng and {Zhang}, Chun-Feng and {Wang}, Wei-Yang and {Wang}, Bo-Jun and {Xu}, Jiang-Wei},
        title = "{FAST Observations of an Extremely Active Episode of FRB 20201124A. IV. Spin Period Search}",
      journal = {Research in Astronomy and Astrophysics},
         year = 2022,
        month = dec,
       volume = {22},
       number = {12},
          eid = {124004},
        pages = {124004},
          doi = {10.1088/1674-4527/ac995d},
archivePrefix = {arXiv},
       eprint = {2210.03610},
 primaryClass = {astro-ph.HE},
       adsurl = {https://ui.adsabs.harvard.edu/abs/2022RAA....22l4004N}
}

@ARTICLE{Du2025,
       author = {{Du}, Chen and {Huang}, Yong-Feng and {Geng}, Jin-Jun and {Gao}, Hao-Xuan and {Zhang}, Li and {Deng}, Chen and {Cui}, Lang and {Liao}, Jie and {Jiang}, Peng-Fei and {Zhang}, Liang and {Wang}, Pei and {Hu}, Chen-Ran and {Dong}, Xiao-Fei and {Xu}, Fan and {Li}, Liang and {Zou}, Ze-Cheng and {Kurban}, Abdusattar},
        title = "{A second-scale periodicity in an active repeating fast radio burst source}",
      journal = {arXiv e-prints},
         year = 2025,
        month = mar,
          eid = {arXiv:2503.12013},
        pages = {arXiv:2503.12013},
          doi = {10.48550/arXiv.2503.12013},
archivePrefix = {arXiv},
       eprint = {2503.12013},
 primaryClass = {astro-ph.HE},
       adsurl = {https://ui.adsabs.harvard.edu/abs/2025arXiv250312013D}
}

@ARTICLE{Lan2024,
       author = {{Lan}, Hao-Tian and {Zhao}, Zhen-Yin and {Wei}, Yu-Jia and {Wang}, Fa-Yin},
        title = "{The Physical Origin of the Periodic Activity for FRB 20180916B}",
      journal = {\apjl},
         year = 2024,
        month = jun,
       volume = {967},
       number = {2},
          eid = {L44},
        pages = {L44},
          doi = {10.3847/2041-8213/ad4ae8},
archivePrefix = {arXiv},
       eprint = {2310.16307},
 primaryClass = {astro-ph.HE},
       adsurl = {https://ui.adsabs.harvard.edu/abs/2024ApJ...967L..44L}
}

@ARTICLE{Li2021,
       author = {{Li}, Dongzi and {Zanazzi}, J.~J.},
        title = "{Emission Properties of Periodic Fast Radio Bursts from the Motion of Magnetars: Testing Dynamical Models}",
      journal = {\apjl},
         year = 2021,
        month = mar,
       volume = {909},
       number = {2},
          eid = {L25},
        pages = {L25},
          doi = {10.3847/2041-8213/abeaa4},
archivePrefix = {arXiv},
       eprint = {2101.05836},
 primaryClass = {astro-ph.HE},
       adsurl = {https://ui.adsabs.harvard.edu/abs/2021ApJ...909L..25L}
}

@ARTICLE{Pastor-Marazuela2021,
       author = {{Pastor-Marazuela}, In{\'e}s and {Connor}, Liam and {van Leeuwen}, Joeri and {Maan}, Yogesh and {ter Veen}, Sander and {Bilous}, Anna and {Oostrum}, Leon and {Petroff}, Emily and {Straal}, Samayra and {Vohl}, Dany and {Attema}, Jisk and {Boersma}, Oliver M. and {Kooistra}, Eric and {van der Schuur}, Daniel and {Sclocco}, Alessio and {Smits}, Roy and {Adams}, Elizabeth A.~K. and {Adebahr}, Bj{\"o}rn and {de Blok}, W.~J.~G. and {Coolen}, Arthur H.~W.~M. and {Damstra}, Sieds and {D{\'e}nes}, Helga and {Hess}, Kelley M. and {van der Hulst}, Thijs and {Hut}, Boudewijn and {Ivashina}, V. Marianna and {Kutkin}, Alexander and {Loose}, G. Marcel and {Lucero}, Danielle M. and {Mika}, {\'A}gnes and {Moss}, Vanessa A. and {Mulder}, Henk and {Norden}, Menno J. and {Oosterloo}, Tom and {Orr{\'u}}, Emanuela and {Ruiter}, Mark and {Wijnholds}, Stefan J.},
        title = "{Chromatic periodic activity down to 120 megahertz in a fast radio burst}",
      journal = {\nat},
         year = 2021,
        month = aug,
       volume = {596},
       number = {7873},
        pages = {505-508},
          doi = {10.1038/s41586-021-03724-8},
archivePrefix = {arXiv},
       eprint = {2012.08348},
 primaryClass = {astro-ph.HE},
       adsurl = {https://ui.adsabs.harvard.edu/abs/2021Natur.596..505P}
}

@ARTICLE{Rajwade2020,
       author = {{Rajwade}, K.~M. and {Mickaliger}, M.~B. and {Stappers}, B.~W. and {Morello}, V. and {Agarwal}, D. and {Bassa}, C.~G. and {Breton}, R.~P. and {Caleb}, M. and {Karastergiou}, A. and {Keane}, E.~F. and {Lorimer}, D.~R.},
        title = "{Possible periodic activity in the repeating FRB 121102}",
      journal = {\mnras},
         year = 2020,
        month = jul,
       volume = {495},
       number = {4},
        pages = {3551-3558},
          doi = {10.1093/mnras/staa1237},
archivePrefix = {arXiv},
       eprint = {2003.03596},
 primaryClass = {astro-ph.HE},
       adsurl = {https://ui.adsabs.harvard.edu/abs/2020MNRAS.495.3551R}
}

@ARTICLE{Cruces2021,
       author = {{Cruces}, M. and {Spitler}, L.~G. and {Scholz}, P. and {Lynch}, R. and {Seymour}, A. and {Hessels}, J.~W.~T. and {Gouiff{\'e}s}, C. and {Hilmarsson}, G.~H. and {Kramer}, M. and {Munjal}, S.},
        title = "{Repeating behaviour of FRB 121102: periodicity, waiting times, and energy distribution}",
      journal = {\mnras},
         year = 2021,
        month = jan,
       volume = {500},
       number = {1},
        pages = {448-463},
          doi = {10.1093/mnras/staa3223},
archivePrefix = {arXiv},
       eprint = {2008.03461},
 primaryClass = {astro-ph.HE},
       adsurl = {https://ui.adsabs.harvard.edu/abs/2021MNRAS.500..448C}
}

@ARTICLE{Chime/FrbCollaboration2020,
       author = {{Chime/Frb Collaboration} and {Amiri}, M. and {Andersen}, B.~C. and {Bandura}, K.~M. and {Bhardwaj}, M. and {Boyle}, P.~J. and {Brar}, C. and {Chawla}, P. and {Chen}, T. and {Cliche}, J.~F. and {Cubranic}, D. and {Deng}, M. and {Denman}, N.~T. and {Dobbs}, M. and {Dong}, F.~Q. and {Fandino}, M. and {Fonseca}, E. and {Gaensler}, B.~M. and {Giri}, U. and {Good}, D.~C. and {Halpern}, M. and {Hessels}, J.~W.~T. and {Hill}, A.~S. and {H{\"o}fer}, C. and {Josephy}, A. and {Kania}, J.~W. and {Karuppusamy}, R. and {Kaspi}, V.~M. and {Keimpema}, A. and {Kirsten}, F. and {Landecker}, T.~L. and {Lang}, D.~A. and {Leung}, C. and {Li}, D.~Z. and {Lin}, H. -H. and {Marcote}, B. and {Masui}, K.~W. and {McKinven}, R. and {Mena-Parra}, J. and {Merryfield}, M. and {Michilli}, D. and {Milutinovic}, N. and {Mirhosseini}, A. and {Naidu}, A. and {Newburgh}, L.~B. and {Ng}, C. and {Nimmo}, K. and {Paragi}, Z. and {Patel}, C. and {Pen}, U. -L. and {Pinsonneault-Marotte}, T. and {Pleunis}, Z. and {Rafiei-Ravandi}, M. and {Rahman}, M. and {Ransom}, S.~M. and {Renard}, A. and {Sanghavi}, P. and {Scholz}, P. and {Shaw}, J.~R. and {Shin}, K. and {Siegel}, S.~R. and {Singh}, S. and {Smegal}, R.~J. and {Smith}, K.~M. and {Stairs}, I.~H. and {Tendulkar}, S.~P. and {Tretyakov}, I. and {Vanderlinde}, K. and {Wang}, H. and {Wang}, X. and {Wulf}, D. and {Yadav}, P. and {Zwaniga}, A.~V.},
        title = "{Periodic activity from a fast radio burst source}",
      journal = {\nat},
         year = 2020,
        month = jun,
       volume = {582},
       number = {7812},
        pages = {351-355},
          doi = {10.1038/s41586-020-2398-2},
archivePrefix = {arXiv},
       eprint = {2001.10275},
 primaryClass = {astro-ph.HE},
       adsurl = {https://ui.adsabs.harvard.edu/abs/2020Natur.582..351C}
}

@ARTICLE{Luo2025,
       author = {{Luo}, Jia-Wei and {Niu}, Jia-Rui and {Wang}, Wei-Yang and {Zhang}, Yong-Kun and {De-Jiang Zhou} and {Xu}, Heng and {Wang}, Pei and {Niu}, Chen-Hui and {Zhang}, Zhen-Hui and {Zhang}, Shuai and {Cai}, Ce and {Han}, Jin-Lin and {Li}, Di and {Lee}, Ke-Jia and {Zhu}, Wei-Wei and {Zhang}, Bing},
        title = "{Hyperactive Repeating Fast Radio Bursts from Rotation-modulated Starquakes on Magnetars}",
      journal = {\apj},
         year = 2025,
        month = jul,
       volume = {988},
       number = {1},
          eid = {62},
        pages = {62},
          doi = {10.3847/1538-4357/ade0b9},
archivePrefix = {arXiv},
       eprint = {2502.16626},
 primaryClass = {astro-ph.HE},
       adsurl = {https://ui.adsabs.harvard.edu/abs/2025ApJ...988...62L}
}

@ARTICLE{LiD2021,
       author = {{Li}, D. and {Wang}, P. and {Zhu}, W.~W. and {Zhang}, B. and {Zhang}, X.~X. and {Duan}, R. and {Zhang}, Y.~K. and {Feng}, Y. and {Tang}, N.~Y. and {Chatterjee}, S. and {Cordes}, J.~M. and {Cruces}, M. and {Dai}, S. and {Gajjar}, V. and {Hobbs}, G. and {Jin}, C. and {Kramer}, M. and {Lorimer}, D.~R. and {Miao}, C.~C. and {Niu}, C.~H. and {Niu}, J.~R. and {Pan}, Z.~C. and {Qian}, L. and {Spitler}, L. and {Werthimer}, D. and {Zhang}, G.~Q. and {Wang}, F.~Y. and {Xie}, X.~Y. and {Yue}, Y.~L. and {Zhang}, L. and {Zhi}, Q.~J. and {Zhu}, Y.},
        title = "{A bimodal burst energy distribution of a repeating fast radio burst source}",
      journal = {\nat},
         year = 2021,
        month = oct,
       volume = {598},
       number = {7880},
        pages = {267-271},
          doi = {10.1038/s41586-021-03878-5},
archivePrefix = {arXiv},
       eprint = {2107.08205},
 primaryClass = {astro-ph.HE},
       adsurl = {https://ui.adsabs.harvard.edu/abs/2021Natur.598..267L}
}

@ARTICLE{Xu2022,
       author = {{Xu}, H. and {Niu}, J.~R. and {Chen}, P. and {Lee}, K.~J. and {Zhu}, W.~W. and {Dong}, S. and {Zhang}, B. and {Jiang}, J.~C. and {Wang}, B.~J. and {Xu}, J.~W. and {Zhang}, C.~F. and {Fu}, H. and {Filippenko}, A.~V. and {Peng}, E.~W. and {Zhou}, D.~J. and {Zhang}, Y.~K. and {Wang}, P. and {Feng}, Y. and {Li}, Y. and {Brink}, T.~G. and {Li}, D.~Z. and {Lu}, W. and {Yang}, Y.~P. and {Caballero}, R.~N. and {Cai}, C. and {Chen}, M.~Z. and {Dai}, Z.~G. and {Djorgovski}, S.~G. and {Esamdin}, A. and {Gan}, H.~Q. and {Guhathakurta}, P. and {Han}, J.~L. and {Hao}, L.~F. and {Huang}, Y.~X. and {Jiang}, P. and {Li}, C.~K. and {Li}, D. and {Li}, H. and {Li}, X.~Q. and {Li}, Z.~X. and {Liu}, Z.~Y. and {Luo}, R. and {Men}, Y.~P. and {Niu}, C.~H. and {Peng}, W.~X. and {Qian}, L. and {Song}, L.~M. and {Stern}, D. and {Stockton}, A. and {Sun}, J.~H. and {Wang}, F.~Y. and {Wang}, M. and {Wang}, N. and {Wang}, W.~Y. and {Wu}, X.~F. and {Xiao}, S. and {Xiong}, S.~L. and {Xu}, Y.~H. and {Xu}, R.~X. and {Yang}, J. and {Yang}, X. and {Yao}, R. and {Yi}, Q.~B. and {Yue}, Y.~L. and {Yu}, D.~J. and {Yu}, W.~F. and {Yuan}, J.~P. and {Zhang}, B.~B. and {Zhang}, S.~B. and {Zhang}, S.~N. and {Zhao}, Y. and {Zheng}, W.~K. and {Zhu}, Y. and {Zou}, J.~H.},
        title = "{A fast radio burst source at a complex magnetized site in a barred galaxy}",
      journal = {\nat},
         year = 2022,
        month = sep,
       volume = {609},
       number = {7928},
        pages = {685-688},
          doi = {10.1038/s41586-022-05071-8},
archivePrefix = {arXiv},
       eprint = {2111.11764},
 primaryClass = {astro-ph.HE},
       adsurl = {https://ui.adsabs.harvard.edu/abs/2022Natur.609..685X}
}

@ARTICLE{Zhou2022,
       author = {{Zhou}, D.~J. and {Han}, J.~L. and {Zhang}, B. and {Lee}, K.~J. and {Zhu}, W.~W. and {Li}, D. and {Jing}, W.~C. and {Wang}, W. -Y. and {Zhang}, Y.~K. and {Jiang}, J.~C. and {Niu}, J.~R. and {Luo}, R. and {Xu}, H. and {Zhang}, C.~F. and {Wang}, B.~J. and {Xu}, J.~W. and {Wang}, P. and {Yang}, Z.~L. and {Feng}, Y.},
        title = "{FAST Observations of an Extremely Active Episode of FRB 20201124A: I. Burst Morphology}",
      journal = {Research in Astronomy and Astrophysics},
         year = 2022,
        month = dec,
       volume = {22},
       number = {12},
          eid = {124001},
        pages = {124001},
          doi = {10.1088/1674-4527/ac98f8},
archivePrefix = {arXiv},
       eprint = {2210.03607},
 primaryClass = {astro-ph.HE},
       adsurl = {https://ui.adsabs.harvard.edu/abs/2022RAA....22l4001Z}
}

@ARTICLE{ZhangYK2023,
       author = {{Zhang}, Yong-Kun and {Li}, Di and {Zhang}, Bing and {Cao}, Shuo and {Feng}, Yi and {Wang}, Wei-Yang and {Qu}, Yuanhong and {Niu}, Jia-Rui and {Zhu}, Wei-Wei and {Han}, Jin-Lin and {Jiang}, Peng and {Lee}, Ke-Jia and {Li}, Dong-Zi and {Luo}, Rui and {Niu}, Chen-Hui and {Tsai}, Chao-Wei and {Wang}, Pei and {Wang}, Fa-Yin and {Wu}, Zi-Wei and {Xu}, Heng and {Yang}, Yuan-Pei and {Zhang}, Jun-Shuo and {Zhou}, De-Jiang and {Zhu}, Yu-Hao},
        title = "{FAST Observations of FRB 20220912A: Burst Properties and Polarization Characteristics}",
      journal = {\apj},
         year = 2023,
        month = oct,
       volume = {955},
       number = {2},
          eid = {142},
        pages = {142},
          doi = {10.3847/1538-4357/aced0b},
archivePrefix = {arXiv},
       eprint = {2304.14665},
 primaryClass = {astro-ph.HE},
       adsurl = {https://ui.adsabs.harvard.edu/abs/2023ApJ...955..142Z}
}

@ARTICLE{Liu2025,
       author = {{Liu}, Xiaohui and {Xu}, Heng and {Niu}, Jiarui and {Zhang}, Yongkun and {Jiang}, Jinchen and {Zhou}, Dejiang and {Han}, Jinlin and {Zhu}, Weiwei and {Lee}, Kejia and {Li}, Di and {Wang}, Wei-Yang and {Zhang}, Bing and {Chen}, Xuelei and {Luo}, Jia-Wei and {Luo}, Rui and {Niu}, Chenhui and {Qu}, Yuanhong and {Wang}, Bojun and {Wang}, Fayin and {Wang}, Pei and {Wang}, Tiancong and {Wu}, Qin and {Wu}, Ziwei and {Xu}, Jiangwei and {Yang}, Yuan-Pei and {Zhang}, Jun-Shuo},
        title = "{Polarization Position Angle Swing and the Rotating Vector Model of Repeating Fast Radio Bursts}",
      journal = {arXiv e-prints},
         year = 2025,
        month = apr,
          eid = {arXiv:2504.00391},
        pages = {arXiv:2504.00391},
          doi = {10.48550/arXiv.2504.00391},
archivePrefix = {arXiv},
       eprint = {2504.00391},
 primaryClass = {astro-ph.HE},
       adsurl = {https://ui.adsabs.harvard.edu/abs/2025arXiv250400391L}
}

@ARTICLE{Zhou2025,
       author = {{Zhou}, Dengke and {Wang}, Pei and {Fang}, Jianhua and {Zhu}, Weiwei and {Zhang}, Bing and {Li}, Di and {Feng}, Yi and {Huang}, Yongfeng and {Lee}, Kejia and {Han}, Jinlin and {Zou}, Yuanchuan and {Zhang}, Junshuo and {Luo}, Rui and {Zhang}, Longxuan and {Wang}, Tiancong and {Lu}, Wanjin and {Cao}, Jinhuang and {Yu}, Wenfei and {Li}, Bing and {Miao}, Chenchen and {Xie}, Jintao and {Chen}, Yunchuan and {Qu}, Yuanhong and {Chen}, Huaxi and {Zhu}, Yuhao and {Cao}, Shuo and {Chen}, Xianglei and {Du}, Chen and {Gao}, He and {Huang}, Yuxiang and {Li}, Ye and {Li}, Jian and {Li}, Dongzi and {Lin}, Lin and {Liu}, Xiaohui and {Luo}, Jiawei and {Niu}, Jiarui and {Niu}, Chenhui and {Qu}, Qingyue and {Tian}, Shiyan and {Tsai}, Chao-Wei and {Wang}, Fayin and {Wang}, Yidan and {Wang}, Weiyang and {Wang}, Bojun and {Weng}, Suming and {Wu}, Qin and {Wu}, Ziwei and {Xu}, Heng and {Yang}, Ai Yuan and {Yew}, Shihan and {Zhang}, Yongkun and {Zhang}, Lei and {Zhang}, Chunfeng and {Zhao}, Rushuang and {Zhou}, Dejiang},
        title = "{Searching for periodic signals and quasi-periodic oscillations from an extremely active cycle of FRB 20240114A}",
      journal = {arXiv e-prints},
         year = 2025,
        month = jul,
          eid = {arXiv:2507.14708},
        pages = {arXiv:2507.14708},
archivePrefix = {arXiv},
       eprint = {2507.14708},
 primaryClass = {astro-ph.HE},
       adsurl = {https://ui.adsabs.harvard.edu/abs/2025arXiv250714708Z}
}

@ARTICLE{Zhang2025,
       author = {{Zhang}, Jun-Shuo and {Wang}, Tian-Cong and {Wang}, Pei and {Gao}, He and {Wu}, Qin and {Li}, Di and {Zhu}, Weiwei and {Zhang}, Bing and {Lee}, Ke-Jia and {Han}, Jinlin and {Tsai}, Chao-Wei and {Wang}, Fayin and {Huang}, Yong-Feng and {Zou}, Yuan-Chuan and {Zhou}, Dengke and {Lu}, Wanjin and {Xie}, Jintao and {Fang}, Jianhua and {Cao}, Jinhuang and {Miao}, Chen-Chen and {Zhu}, Yuhao and {Chen}, Yunchuan and {Cheng}, Xiaofeng and {Ke}, Yinan and {Zhang}, Yong-Kun and {Zhang}, Long-Xuan and {Cao}, Shuo and {Tian}, Shiyan and {Wu}, Zi-Wei and {Zhang}, Chunfeng and {Niu}, Jiarui and {Zhou}, Dejiang and {Xu}, Silu and {Wang}, Bojun and {Chen}, Huaxi and {Chen}, Xiang-Lei and {Cui}, Xianghan and {Feng}, Yi and {G{\"u}gercino{\u{g}}lu}, Erbil and {Huang}, Yu-Xiang and {Li}, Jian and {Li}, Dongming and {Li}, Dong-Zi and {Li}, Ye and {Lin}, Lin and {Liu}, Xiaohui and {Luo}, Rui and {Luo}, Jia-Wei and {Niu}, Chen-Hui and {Qu}, Qingyue and {Qu}, Yuanhong and {Menberu Tedila}, Habtamu and {Wang}, Chengjie and {Wang}, Wei-Yang and {Wang}, Yabiao and {Wang}, Yi-Dan and {Weng}, Suming and {Wu}, Yunsheng and {Xu}, Heng and {Yang}, Aiyuan and {Yang}, Yuan-Pei and {Yew}, Shihan and {Yu}, Wenfei and {Zhang}, Lei and {Zhao}, Rushuang},
        title = "{A prolific repeating fast radio burst source and a crisis of the magnetar model}",
      journal = {arXiv e-prints},
         year = 2025,
        month = jul,
          eid = {arXiv:2507.14707},
        pages = {arXiv:2507.14707},
archivePrefix = {arXiv},
       eprint = {2507.14707},
 primaryClass = {astro-ph.HE},
       adsurl = {https://ui.adsabs.harvard.edu/abs/2025arXiv250714707Z}
}

@ARTICLE{Platts2019,
       author = {{Platts}, E. and {Weltman}, A. and {Walters}, A. and {Tendulkar}, S.~P. and {Gordin}, J.~E.~B. and {Kandhai}, S.},
        title = "{A living theory catalogue for fast radio bursts}",
      journal = {\physrep},
         year = 2019,
        month = aug,
       volume = {821},
        pages = {1-27},
          doi = {10.1016/j.physrep.2019.06.003},
archivePrefix = {arXiv},
       eprint = {1810.05836},
 primaryClass = {astro-ph.HE},
       adsurl = {https://ui.adsabs.harvard.edu/abs/2019PhR...821....1P}
}

@ARTICLE{Kumar2020,
       author = {{Kumar}, Pawan and {Bo{\v{s}}njak}, {\v{Z}}eljka},
        title = "{FRB coherent emission from decay of Alfv{\'e}n waves}",
      journal = {\mnras},
         year = 2020,
        month = may,
       volume = {494},
       number = {2},
        pages = {2385-2395},
          doi = {10.1093/mnras/staa774},
archivePrefix = {arXiv},
       eprint = {2004.00644},
 primaryClass = {astro-ph.HE},
       adsurl = {https://ui.adsabs.harvard.edu/abs/2020MNRAS.494.2385K}
}

@ARTICLE{Liang2025,
       author = {{Liang}, Yi-Fang and {Li}, Ye and {Tang}, Zhen-Fan and {Yang}, Xuan and {Zhang}, Song-Bo and {Yang}, Yuan-Pei and {Wang}, Fa-Yin and {Wang}, Bao and {Xiao}, Di and {Zhao}, Qing and {Wei}, Jun-Jie and {Geng}, Jin-Jun and {Niu}, Jia-Rui and {Zhang}, Jun-Shuo and {Chen}, Guo and {Fang}, Min and {Wu}, Xue-Feng and {Dai}, Zi-Gao and {Zhu}, Wei-Wei and {Jiang}, Peng and {Zhang}, Bing},
        title = "{A possible periodic RM evolution in the repeating FRB 20220529}",
      journal = {arXiv e-prints},
         year = 2025,
        month = may,
          eid = {arXiv:2505.10463},
        pages = {arXiv:2505.10463},
          doi = {10.48550/arXiv.2505.10463},
archivePrefix = {arXiv},
       eprint = {2505.10463},
 primaryClass = {astro-ph.HE},
       adsurl = {https://ui.adsabs.harvard.edu/abs/2025arXiv250510463L}
}

@ARTICLE{Xu2025,
       author = {{Xu}, Jiangwei and {Xu}, Heng and {Guo}, Yanjun and {Jiang}, Jinchen and {Wang}, Bojun and {Xue}, Zihan and {Men}, Yunpeng and {Lee}, Kejia and {Zhang}, Bing and {Zhu}, Weiwei and {Han}, Jinlin},
        title = "{Periodic variation of magnetoionic environment of a fast radio burst source}",
      journal = {arXiv e-prints},
         year = 2025,
        month = may,
          eid = {arXiv:2505.06006},
        pages = {arXiv:2505.06006},
          doi = {10.48550/arXiv.2505.06006},
archivePrefix = {arXiv},
       eprint = {2505.06006},
 primaryClass = {astro-ph.HE},
       adsurl = {https://ui.adsabs.harvard.edu/abs/2025arXiv250506006X}
}

@ARTICLE{Jahns2023,
       author = {{Jahns}, J.~N. and {Spitler}, L.~G. and {Nimmo}, K. and {Hewitt}, D.~M. and {Snelders}, M.~P. and {Seymour}, A. and {Hessels}, J.~W.~T. and {Gourdji}, K. and {Michilli}, D. and {Hilmarsson}, G.~H.},
        title = "{The FRB 20121102A November rain in 2018 observed with the Arecibo Telescope}",
      journal = {\mnras},
         year = 2023,
        month = feb,
       volume = {519},
       number = {1},
        pages = {666-687},
          doi = {10.1093/mnras/stac3446},
archivePrefix = {arXiv},
       eprint = {2202.05705},
 primaryClass = {astro-ph.HE},
       adsurl = {https://ui.adsabs.harvard.edu/abs/2023MNRAS.519..666J}
}

@ARTICLE{Beniamini2020,
       author = {{Beniamini}, Paz and {Wadiasingh}, Zorawar and {Metzger}, Brian D.},
        title = "{Periodicity in recurrent fast radio bursts and the origin of ultralong period magnetars}",
      journal = {\mnras},
         year = 2020,
        month = aug,
       volume = {496},
       number = {3},
        pages = {3390-3401},
          doi = {10.1093/mnras/staa1783},
archivePrefix = {arXiv},
       eprint = {2003.12509},
 primaryClass = {astro-ph.HE},
       adsurl = {https://ui.adsabs.harvard.edu/abs/2020MNRAS.496.3390B}
}

@ARTICLE{Zanazzi2020,
       author = {{Zanazzi}, J.~J. and {Lai}, Dong},
        title = "{Periodic Fast Radio Bursts with Neutron Star Free Precession}",
      journal = {\apjl},
         year = 2020,
        month = mar,
       volume = {892},
       number = {1},
          eid = {L15},
        pages = {L15},
          doi = {10.3847/2041-8213/ab7cdd},
archivePrefix = {arXiv},
       eprint = {2002.05752},
 primaryClass = {astro-ph.HE},
       adsurl = {https://ui.adsabs.harvard.edu/abs/2020ApJ...892L..15Z}
}

@ARTICLE{Levin2020,
       author = {{Levin}, Yuri and {Beloborodov}, Andrei M. and {Bransgrove}, Ashley},
        title = "{Precessing Flaring Magnetar as a Source of Repeating FRB 180916.J0158+65}",
      journal = {\apjl},
         year = 2020,
        month = jun,
       volume = {895},
       number = {2},
          eid = {L30},
        pages = {L30},
          doi = {10.3847/2041-8213/ab8c4c},
archivePrefix = {arXiv},
       eprint = {2002.04595},
 primaryClass = {astro-ph.HE},
       adsurl = {https://ui.adsabs.harvard.edu/abs/2020ApJ...895L..30L}
}

@ARTICLE{Ioka2020,
       author = {{Ioka}, Kunihito and {Zhang}, Bing},
        title = "{A Binary Comb Model for Periodic Fast Radio Bursts}",
      journal = {\apjl},
         year = 2020,
        month = apr,
       volume = {893},
       number = {1},
          eid = {L26},
        pages = {L26},
          doi = {10.3847/2041-8213/ab83fb},
archivePrefix = {arXiv},
       eprint = {2002.08297},
 primaryClass = {astro-ph.HE},
       adsurl = {https://ui.adsabs.harvard.edu/abs/2020ApJ...893L..26I}
}

@ARTICLE{Lyutikov2020,
       author = {{Lyutikov}, Maxim and {Barkov}, Maxim V. and {Giannios}, Dimitrios},
        title = "{FRB Periodicity: Mild Pulsars in Tight O/B-star Binaries}",
      journal = {\apjl},
         year = 2020,
        month = apr,
       volume = {893},
       number = {2},
          eid = {L39},
        pages = {L39},
          doi = {10.3847/2041-8213/ab87a4},
archivePrefix = {arXiv},
       eprint = {2002.01920},
 primaryClass = {astro-ph.HE},
       adsurl = {https://ui.adsabs.harvard.edu/abs/2020ApJ...893L..39L}
}

@ARTICLE{Wada2021,
       author = {{Wada}, Tomoki and {Ioka}, Kunihito and {Zhang}, Bing},
        title = "{Binary Comb Models for FRB 121102}",
      journal = {\apj},
         year = 2021,
        month = oct,
       volume = {920},
       number = {1},
          eid = {54},
        pages = {54},
          doi = {10.3847/1538-4357/ac127a},
archivePrefix = {arXiv},
       eprint = {2105.14480},
 primaryClass = {astro-ph.HE},
       adsurl = {https://ui.adsabs.harvard.edu/abs/2021ApJ...920...54W}
}

@ARTICLE{Yang2020,
       author = {{Yang}, Huan and {Zou}, Yuan-Chuan},
        title = "{Orbit-induced Spin Precession as a Possible Origin for Periodicity in Periodically Repeating Fast Radio Bursts}",
      journal = {\apjl},
         year = 2020,
        month = apr,
       volume = {893},
       number = {2},
          eid = {L31},
        pages = {L31},
          doi = {10.3847/2041-8213/ab800f},
archivePrefix = {arXiv},
       eprint = {2002.02553},
 primaryClass = {astro-ph.HE},
       adsurl = {https://ui.adsabs.harvard.edu/abs/2020ApJ...893L..31Y}
}

@ARTICLE{Bethapudi2025,
       author = {{Bethapudi}, S. and {Li}, D.~Z. and {Spitler}, L.~G. and {Marthi}, V.~R. and {Bause}, M.~L. and {Main}, R.~A. and {Wharton}, R.~S.},
        title = "{Constraining the origin of the long term periodicity of FRB 20180916B with Polarization Position Angle}",
      journal = {arXiv e-prints},
         year = 2025,
        month = jul,
          eid = {arXiv:2507.07651},
        pages = {arXiv:2507.07651},
          doi = {10.48550/arXiv.2507.07651},
archivePrefix = {arXiv},
       eprint = {2507.07651},
 primaryClass = {astro-ph.HE},
       adsurl = {https://ui.adsabs.harvard.edu/abs/2025arXiv250707651B}
}

@ARTICLE{Katz2021,
       author = {{Katz}, J.~I.},
        title = "{Testing models of periodically modulated FRB activity}",
      journal = {\mnras},
         year = 2021,
        month = apr,
       volume = {502},
       number = {3},
        pages = {4664-4668},
          doi = {10.1093/mnras/stab399},
archivePrefix = {arXiv},
       eprint = {2012.15354},
 primaryClass = {astro-ph.HE},
       adsurl = {https://ui.adsabs.harvard.edu/abs/2021MNRAS.502.4664K}
}

@ARTICLE{Wang2022,
       author = {{Wang}, F.~Y. and {Zhang}, G.~Q. and {Dai}, Z.~G. and {Cheng}, K.~S.},
        title = "{Repeating fast radio burst 20201124A originates from a magnetar/Be star binary}",
      journal = {Nature Communications},
         year = 2022,
        month = sep,
       volume = {13},
          eid = {4382},
        pages = {4382},
          doi = {10.1038/s41467-022-31923-y},
archivePrefix = {arXiv},
       eprint = {2204.08124},
 primaryClass = {astro-ph.HE},
       adsurl = {https://ui.adsabs.harvard.edu/abs/2022NatCo..13.4382W}
}

@ARTICLE{Caleb2020,
       author = {{Caleb}, M. and {Stappers}, B.~W. and {Abbott}, T.~D. and {Barr}, E.~D. and {Bezuidenhout}, M.~C. and {Buchner}, S.~J. and {Burgay}, M. and {Chen}, W. and {Cognard}, I. and {Driessen}, L.~N. and {Fender}, R. and {Hilmarsson}, G.~H. and {Hoang}, J. and {Horn}, D.~M. and {Jankowski}, F. and {Kramer}, M. and {Lorimer}, D.~R. and {Malenta}, M. and {Morello}, V. and {Pilia}, M. and {Platts}, E. and {Possenti}, A. and {Rajwade}, K.~M. and {Ridolfi}, A. and {Rhodes}, L. and {Sanidas}, S. and {Serylak}, M. and {Spitler}, L.~G. and {Townsend}, L.~J. and {Weltman}, A. and {Woudt}, P.~A. and {Wu}, J.},
        title = "{Simultaneous multi-telescope observations of FRB 121102}",
      journal = {\mnras},
         year = 2020,
        month = aug,
       volume = {496},
       number = {4},
        pages = {4565-4573},
          doi = {10.1093/mnras/staa1791},
archivePrefix = {arXiv},
       eprint = {2006.08662},
 primaryClass = {astro-ph.HE},
       adsurl = {https://ui.adsabs.harvard.edu/abs/2020MNRAS.496.4565C}
}

@ARTICLE{WangP2025,
       author = {{Wang}, P. and {Zhang}, J.~S. and {Yang}, Y.~P. and {Zhou}, D.~K. and {Zhang}, Y.~K. and {Feng}, Y. and {Zhao}, Z.~Y. and {Fang}, J.~H. and {Li}, D. and {Zhu}, W.~W. and {Zhang}, B. and {Wang}, F.~Y. and {Huang}, Y.~F. and {Luo}, R. and {Han}, J.~L. and {Lee}, K.~J. and {Tsai}, C.~W. and {Dai}, Z.~G. and {Gao}, H. and {Zheng}, X.~P. and {Cao}, J.~H. and {Chen}, X.~L. and {Gugercinoglu}, E. and {Jiang}, J.~C. and {Jing}, W.~C. and {Li}, Y. and {Li}, J. and {Lu}, W.~J. and {Luo}, J.~W. and {Lyu}, F. and {Miao}, C.~C. and {Niu}, C.~H. and {Niu}, J.~R. and {Qu}, Y. and {Wang}, W.~Y. and {Wang}, Y.~D. and {Wang}, Y.~B. and {Wang}, C.~J. and {Wu}, Q. and {Wu}, Y.~S. and {Weng}, S.~M. and {Xiao}, D. and {Xu}, H. and {Yao}, J.~M. and {Zhang}, C.~F. and {Zhao}, R.~S. and {Liu}, Q.~Z. and {Zhang}, J. and {Zhou}, D.~J. and {Zhang}, L. and {Zhu}, Y.~H.},
        title = "{Decadal evolution of a repeating fast radio burst source}",
      journal = {arXiv e-prints},
         year = 2025,
        month = jul,
          eid = {arXiv:2507.15790},
        pages = {arXiv:2507.15790},
          doi = {10.48550/arXiv.2507.15790},
archivePrefix = {arXiv},
       eprint = {2507.15790},
 primaryClass = {astro-ph.HE},
       adsurl = {https://ui.adsabs.harvard.edu/abs/2025arXiv250715790W}
}

@ARTICLE{Morello2020,
       author = {{Morello}, V. and {Barr}, E.~D. and {Stappers}, B.~W. and {Keane}, E.~F. and {Lyne}, A.~G.},
        title = "{Optimal periodicity searching: revisiting the fast folding algorithm for large-scale pulsar surveys}",
      journal = {\mnras},
         year = 2020,
        month = oct,
       volume = {497},
       number = {4},
        pages = {4654-4671},
          doi = {10.1093/mnras/staa2291},
archivePrefix = {arXiv},
       eprint = {2004.03701},
 primaryClass = {astro-ph.IM},
       adsurl = {https://ui.adsabs.harvard.edu/abs/2020MNRAS.497.4654M}
}

@software{Ransom2011,
       author = {{Ransom}, Scott},
        title = "{PRESTO: PulsaR Exploration and Search TOolkit}",
 howpublished = {Astrophysics Source Code Library, record ascl:1107.017},
         year = 2011,
        month = jul,
          eid = {ascl:1107.017},
archivePrefix = {ascl},
       eprint = {1107.017},
       adsurl = {https://ui.adsabs.harvard.edu/abs/2011ascl.soft07017R}
}

@ARTICLE{Espinoza-Dupouy2025,
       author = {{Espinoza-Dupouy}, M.~C. and {Cruces}, M. and {Cassanelli}, T. and {Braga}, C.~A. and {Berm{\'u}dez}, E. and {Vera-Casanova}, J.},
        title = "{Chromatic activity window of periodic FRBs: FRB 20121102A and FRB 20180916B}",
      journal = {arXiv e-prints},
         year = 2025,
        month = jul,
          eid = {arXiv:2507.04609},
        pages = {arXiv:2507.04609},
          doi = {10.48550/arXiv.2507.04609},
archivePrefix = {arXiv},
       eprint = {2507.04609},
 primaryClass = {astro-ph.HE},
       adsurl = {https://ui.adsabs.harvard.edu/abs/2025arXiv250704609E}
}

@ARTICLE{VanderPlas2018,
       author = {{VanderPlas}, Jacob T.},
        title = "{Understanding the Lomb-Scargle Periodogram}",
      journal = {\apjs},
         year = 2018,
        month = may,
       volume = {236},
       number = {1},
          eid = {16},
        pages = {16},
          doi = {10.3847/1538-4365/aab766},
archivePrefix = {arXiv},
       eprint = {1703.09824},
 primaryClass = {astro-ph.IM},
       adsurl = {https://ui.adsabs.harvard.edu/abs/2018ApJS..236...16V}
}

@ARTICLE{Lomb1976,
       author = {{Lomb}, N.~R.},
        title = "{Least-Squares Frequency Analysis of Unequally Spaced Data}",
      journal = {\apss},
         year = 1976,
        month = feb,
       volume = {39},
       number = {2},
        pages = {447-462},
          doi = {10.1007/BF00648343},
       adsurl = {https://ui.adsabs.harvard.edu/abs/1976Ap&SS..39..447L}
}

@ARTICLE{Scargle1982,
       author = {{Scargle}, J.~D.},
        title = "{Studies in astronomical time series analysis. II. Statistical aspects of spectral analysis of unevenly spaced data.}",
      journal = {\apj},
         year = 1982,
        month = dec,
       volume = {263},
        pages = {835-853},
          doi = {10.1086/160554},
       adsurl = {https://ui.adsabs.harvard.edu/abs/1982ApJ...263..835S}
}

@ARTICLE{Buccheri1983,
       author = {{Buccheri}, R. and {Bennett}, K. and {Bignami}, G.~F. and {Bloemen}, J.~B.~G.~M. and {Boriakoff}, V. and {Caraveo}, P.~A. and {Hermsen}, W. and {Kanbach}, G. and {Manchester}, R.~N. and {Masnou}, J.~L. and {Mayer-Hasselwander}, H.~A. and {{\"O}zel}, M.~E. and {Paul}, J.~A. and {Sacco}, B. and {Scarsi}, L. and {Strong}, A.~W.},
        title = "{Search for pulsed {\ensuremath{\gamma}}-ray emission from radio pulsars in the COS-B data.}",
      journal = {\aap},
         year = 1983,
        month = dec,
       volume = {128},
        pages = {245-251},
       adsurl = {https://ui.adsabs.harvard.edu/abs/1983A&A...128..245B}
}

@ARTICLE{1989A&A...221..180D,
       author = {{de Jager}, O.~C. and {Raubenheimer}, B.~C. and {Swanepoel}, J.~W.~H.},
        title = "{A powerful test for weak periodic signals with unknown light curve shape in sparse data.}",
      journal = {\aap},
         year = 1989,
        month = aug,
       volume = {221},
        pages = {180-190},
       adsurl = {https://ui.adsabs.harvard.edu/abs/1989A&A...221..180D}
}

@ARTICLE{Leahy1983,
       author = {{Leahy}, D.~A. and {Elsner}, R.~F. and {Weisskopf}, M.~C.},
        title = "{On searches for periodic pulsed emission - The Rayleigh test compared to epoch folding}",
      journal = {\apj},
         year = 1983,
        month = sep,
       volume = {272},
        pages = {256-258},
          doi = {10.1086/161288},
       adsurl = {https://ui.adsabs.harvard.edu/abs/1983ApJ...272..256L}
}

@ARTICLE{Protheroe1987,
       author = {{Protheroe}, R.~J.},
        title = "{Periodic analysis of gamma-ray data.}",
      journal = {\pasa},
         year = 1987,
        month = jan,
       volume = {7},
       number = {2},
        pages = {167-172},
          doi = {10.1017/S1323358000022153},
       adsurl = {https://ui.adsabs.harvard.edu/abs/1987PASA....7..167P}
}

@ARTICLE{Swanepoel1996,
       author = {{Swanepoel}, J.~W.~H. and {de Beer}, C.~F. and {Loots}, H.},
        title = "{Estimation of the Strength of a Periodic Signal from Photon Arrival Times}",
      journal = {\apj},
         year = 1996,
        month = aug,
       volume = {467},
        pages = {261},
          doi = {10.1086/177600},
       adsurl = {https://ui.adsabs.harvard.edu/abs/1996ApJ...467..261S}
}

@ARTICLE{ZhangML2018,
       author = {{Zhang}, Yunfan Gerry and {Gajjar}, Vishal and {Foster}, Griffin and {Siemion}, Andrew and {Cordes}, James and {Law}, Casey and {Wang}, Yu},
        title = "{Fast Radio Burst 121102 Pulse Detection and Periodicity: A Machine Learning Approach}",
      journal = {\apj},
         year = 2018,
        month = oct,
       volume = {866},
       number = {2},
          eid = {149},
        pages = {149},
          doi = {10.3847/1538-4357/aadf31},
archivePrefix = {arXiv},
       eprint = {1809.03043},
 primaryClass = {astro-ph.HE},
       adsurl = {https://ui.adsabs.harvard.edu/abs/2018ApJ...866..149Z}
}

@BOOK{Cox1962,
       author = {{Cox}, David R.},
        title = "{Renewal Theory}",
         year = 1962,
    publisher = {Methuen},
      address = {London},
       series = {Methuen's Monographs on Applied Probability and Statistics},
        pages = {142}
}

@BOOK{DaleyVereJones2003,
       author = {{Daley}, D. J. and {Vere-Jones}, D.},
        title = "{An Introduction to the Theory of Point Processes. Volume I: Elementary Theory and Methods}",
         year = 2003,
    publisher = {Springer},
      address = {New York},
       series = {Probability and Its Applications},
      edition = {2},
        pages = {471},
          doi = {10.1007/b97277},
         isbn = {978-0-387-95541-4}
}

@BOOK{Asmussen2003,
       author = {{Asmussen}, S{\o}ren},
        title = "{Applied Probability and Queues}",
         year = 2003,
    publisher = {Springer},
      address = {New York},
       series = {Stochastic Modelling and Applied Probability},
       volume = {51},
      edition = {2},
        pages = {438},
          doi = {10.1007/b97236},
         isbn = {978-0-387-00211-8}
}

@ARTICLE{Ogata1988,
       author = {{Ogata}, Yosihiko},
        title = "{Statistical Models for Earthquake Occurrences and Residual Analysis for Point Processes}",
      journal = {Journal of the American Statistical Association},
         year = 1988,
       volume = {83},
       number = {401},
        pages = {9-27},
          doi = {10.1080/01621459.1988.10478560}
}

@ARTICLE{Ogata1998,
       author = {{Ogata}, Yosihiko},
        title = "{Space-Time Point-Process Models for Earthquake Occurrences}",
      journal = {Annals of the Institute of Statistical Mathematics},
         year = 1998,
        month = jun,
       volume = {50},
       number = {2},
        pages = {379-402},
          doi = {10.1023/A:1003403601725},
          url = {https://doi.org/10.1023/A:1003403601725}
}

@ARTICLE{Brown2002,
       author = {{Brown}, Emery N. and {Barbieri}, Riccardo and {Ventura}, Val{\'e}rie and {Kass}, Robert E. and {Frank}, Loren M.},
        title = "{The Time-Rescaling Theorem and Its Application to Neural Spike Train Data Analysis}",
      journal = {Neural Computation},
         year = 2002,
        month = feb,
       volume = {14},
       number = {2},
        pages = {325-346},
          doi = {10.1162/08997660252741149}
}

@ARTICLE{Truccolo2005,
       author = {{Truccolo}, Wilson and {Eden}, Uri T. and {Fellows}, Matthew R. and {Donoghue}, John P. and {Brown}, Emery N.},
        title = "{A Point Process Framework for Relating Neural Spiking Activity to Spiking History, Neural Ensemble, and Extrinsic Covariate Effects}",
      journal = {Journal of Neurophysiology},
         year = 2005,
        month = feb,
       volume = {93},
       number = {2},
        pages = {1074-1089},
          doi = {10.1152/jn.00697.2004}
}

%\begin{figure}
%	\centering
%	\includegraphics[width=0.8\columnwidth]{hist_Duration.pdf}
%	\caption{The duration distribution of the sliding windows and the ranodom selected burst set for FRB 20240114A. The left panel shows The duration distribution of the sliding windows and the right panel shows the distribution of the burst windows derived by random selected.}\label{fig_duration}
%\end{figure}

%\begin{figure}
%	\centering
%	\includegraphics[width=0.8\columnwidth]{hist_all_FRB_HZ_pf.pdf}
%	\caption{The distribution of all period candidates for four active repeating FRB sources with use of two different methods. The x-axis represents the period in seconds on a logarithmic scale, ranging from $0.01$ s to $100$ s, while the y-axis indicates the number of detected period candidates in each bin. The blue histogram represents the results derived from the $Z_{m}^2$ method, while the orange histogram represents the results derived from the phase-folding method.}\label{fig_HZ_pf}
%\end{figure}

\begin{table}
\label{table1}
\caption{Best-fit parameters of the rotation-modulated ETAS model for FRB 20240114A.}
\label{tab:etas_parameters}
\begin{center}
\begin{tabular}{lcc}
\hline
Parameter & Value & Description \\
\hline
$\alpha$ & 2.95 & Modulation phase parameter (deg) \\
$\beta$ & $-0.31$ & Modulation phase parameter (deg) \\
$P$ & 1.84 & Rotation period parameter (s) \\
$\Gamma$ & 146.33 & Rotation-modulation strength \\
$\rho_{\rm max}$ & 9.80 & Maximum modulation angle (deg) \\
$\log_{10}\mu$ & $-1.08$ & Background burst rate parameter \\
$\log_{10}K$ & $-3.26$ & Burst productivity parameter \\
$a$ & $-0.44$ & Magnitude productivity scaling parameter \\
$\log_{10}c$ & $-0.68$ & Omori time offset parameter \\
$p$ & 5.85 & Temporal decay index \\
$\beta_e$ & 2.29 & Energy distribution index \\
$m_c$ & 39.42 & Critical magnitude threshold \\
\hline
\end{tabular}
\end{center}
\end{table}
	
\end{document}